\RequirePackage{fix-cm}
\documentclass[twocolumn]{svjour3}       % onecolumn (second format)
\smartqed  % flush right qed marks, e.g. at end of proof
\usepackage{graphicx, hyperref, breakurl}
\usepackage{epstopdf, bbm}
\usepackage{natbib, url ,amsmath, amssymb, latexsym, appendix}
\usepackage{float, multirow, subcaption, enumitem} 
\floatstyle{plain}
\newfloat{Algorithm}{thp}{lop}
\floatname{Algorithm}{Algorithm}

\newcommand{\ra}[1]{\renewcommand{\arraystretch}{#1}}
\def\e{{\rm e}}                        % natural e
\allowdisplaybreaks
\graphicspath{{images/}}

\begin{document}
\sloppy

\title{Gaussian variational approximation with sparse precision matrices}


\author{Linda S. L. Tan  \and  David J. Nott }


\institute{Linda S. L. Tan \at
	Department of Statistics and Applied Probability \\
	National University of Singapore \\
	6 Science Drive 2, Singapore 117546 \\
	Tel.: +65-6516-4416\\ 
	Fax: +65-6872-3919\\  
	\email{statsll@nus.edu.sg}           
	\and
	David J. Nott \at
	Department of Statistics and Applied Probability \\
	National University of Singapore \\
	6 Science Drive 2, Singapore 117546 \\
	Operations Research and Analytics Cluster \\
	National University of Singapore \\
	21 Lower Kent Ridge Road, Singapore 119077 \\	
	Tel.: +65-6516-2744\\ 
	Fax: +65-6872-3919\\  
	\email{standj@nus.edu.sg} 
}

\date{Received: date / Accepted: date}

\maketitle

\begin{abstract} % 199 words
We consider the problem of learning a Gaussian variational approximation to the posterior distribution for a high-dimensional parameter,
where we impose sparsity in the precision matrix to reflect appropriate conditional independence structure in the model. Incorporating sparsity in the precision matrix allows the Gaussian variational distribution to be both flexible and parsimonious, and the sparsity is achieved through parameterization in terms of the Cholesky factor. Efficient stochastic gradient methods which make appropriate use of gradient information for the target distribution are developed for the optimization. We consider alternative estimators of the stochastic gradients which have lower variation and are more stable. Our approach is illustrated using generalized linear mixed models and state space models for time series.

\end{abstract}

\section{Introduction}\label{sec:Intro}
Bayesian inference provides a principled way of combining data with prior beliefs through the application of Bayes' rule. The posterior distribution is, however, often intractable and simulation-based Markov Chain Monte Carlo (MCMC) methods have become a central tool in Bayesian computation. In recent years, variational methods \citep{Jordan1999} have also emerged as an important alternative to MCMC, providing fast  approximate inference for complex, high-dimensional models. Unlike MCMC, which can be made arbitrarily accurate, variational methods make certain simplifying assumptions about the posterior density (e.g. a tractable form $q(\theta)$ where $\theta$ denotes the vector of variables) and seek to optimize the Kullback-Leibler divergence $D_{KL}(q||p)$ between $q(\theta)$ and the true posterior $p(\theta|y)$ subject to these assumed restrictions. While earlier research on variational methods concentrated on conjugate models with analytically tractable expectations under which the variational Bayes approach \citep{Attias1999} yields efficient closed-form updates \citep{Winn2005}, recent focus considers stochastic gradient approximation methods \citep{Robbins1951} for non-conjugate models \citep[e.g.][]{Paisley2012, Salimans2013}. Further discussion of the literature is deferred to Section \ref{sec:TL}. \cite{Rohde2015} give a nice recent summary of alternatives to stochastic gradient approaches for handling non-conjugacy in the variational Bayes framework.

\cite{Titsias2014} propose a simple yet effective variational method known as ``doubly stochastic variational inference", where the approximating density is parameterized in terms of its mean $\mu$ and a lower triangular scale matrix $L$. An efficient stochastic gradient algorithm is then developed for optimizing $\mu$ and $L$ by (1) parameterizing the vector of variables $\theta$ as $Lz+\mu$ where $z$ is a random variable that can be sampled easily from a base distribution that does not depend on the variational parameters \citep[see also][]{Kingma2014, Rezende2014} and (2) sub-sampling from the data. The stochastic gradients constructed in this manner are ``doubly stochastic" as they are built upon two sources of stochasticity that comes from sampling from the variational distribution and the full data set. This approach is very general in that it can be applied to any model where the joint density is differentiable. Unlike variational Bayes, it does not assume independence relationships among blocks of an appropriate partition of $\theta$. Such independence assumptions have been shown to result in underestimation of the posterior variance \citep{Wang2005, Bishop2006}. The quality of the resulting approximation is thus limited only by how well the form of $q(\theta)$ matches the true posterior. Using this approach, \cite{Kucukelbir2016} develop an automatic differentiation variational inference (ADVI) algorithm in Stan, where $q(\theta)$ is assumed to be either a diagonal (mean-field) or unrestricted Gaussian variational approximation. Constrained variables are transformed to the real line via Stan's library of transformations and the gradients are computed using Monte Carlo integration. They note that while unrestricted ADVI is able to capture posterior correlations and hence produces more accurate marginal variance estimates than mean field ADVI, it can be prohibitively slow for large data since the number of variational parameters scales as the square of the length of $\theta$.

In this article, we consider variational approximations which take the form of a multivariate Gaussian distribution $N(\mu, \Sigma)$ for models with high-dimensional parameters ($\mu$ denotes the mean and $\Sigma$ the covariance matrix). However, instead of expressing $\Sigma$ as $LL^T$ and optimizing $L$ (the Cholesky factor of $\Sigma$) as in \cite{Titsias2014} and \cite{Kucukelbir2016}, we parameterize the optimization problem in terms of the Cholesky factor of the precision matrix. This parameterization is important as it provides an avenue to impose a sparsity structure in the precision matrix that reflects conditional independence relationships in the posterior. This sparsity structure reduces computational complexity greatly and enables fast inference for models with a large number of variables without having to assume independence relationships in the posterior. We demonstrate how our approach can be applied to generalized linear mixed models (GLMMs) and state space models (SSMs) for time series data. Assuming that the number of global variables is small compared to the number of local variables, our approach reduces the number of variational parameters to be updated in each iteration from $\mathcal{O}(n^2)$ to $\mathcal{O}(n)$, where $n$ denotes the number of subjects in GLMMs and the length of time series in SSMs. In this way, the accuracy of using a unrestricted lower triangular matrix $L$ can be achieved at the computational cost (same order of magnitude) of using a diagonal matrix $L$. 

Recently, several classes of richer variational approximations which go beyond factorized (mean-field) approximating densities and which are able to reflect the posterior dependence structure to varying degrees have been proposed \citep[e.g.][]{Gershman2012, Salimans2013}. \cite{Rezende2015} propose the specification of the approximate posterior using normalizing flows. Starting with say simple factorized distributions, highly flexible and complex approximate posteriors are constructed by transforming the initial density through a sequence of invertible mappings which perform expansions or contractions of the probability mass in targeted regions. The resulting chain of transformed densities is known as a normalizing flow. The authors show that true posterior can be recovered asymptotically under the Langevin flow, thus overcoming an important limitation of variational inference. \cite{Ranganath2016} propose hierarchical variational models which are built by placing a prior distribution on the parameters of a mean-field variational approximation and then proceeding to integrate out the mean-field parameters. They specify the prior using normalizing flows and demonstrate that the hierarchical variational model achieves better performance in terms of perplexity and held-out likelihood for deep exponential families \citep{Ranganath2015}. Structured stochastic variational inference \citep{Hoffman2015} is a generalization of stochastic variational inference to allow dependencies between global and local variables. In the approximating density, independence is assumed only among elements in the global variables and among the local variables conditional on the global variables. Dependence between a local variable and the global variables is captured via a local parameter defined implicitly as the point at which the local evidence lower bound is maximized.

\cite{Archer2016} develop ``black-box" variational inference \citep{Ranganath2014} for SSMs, where a Gaussian variational approximation is considered for the latent states. To capture the temporal correlation structure, the precision matrix is assumed to be a block tri-diagonal matrix. While related, our approach differs from \cite{Archer2016} in several aspects. For the SSMs application, we consider a joint Gaussian variational approximation for the model parameters and latent states while \cite{Archer2016} assume that the model parameters are known and consider a Gaussian approximate posterior for the latent states only. Secondly, we optimize the Cholesky factor of the precision matrix directly while \cite{Archer2016} consider other parameterizations such as defining the approximate posterior through a product of Gaussian factors and parameterizing the mean and blocks in the tri-diagonal inverse covariance using neural networks. Third, we consider a more general sparsity structure in the precision matrix, which reflects the conditional independence structures in the posterior distribution and is not limited to band matrices. We also consider an alternative estimator of the stochastic gradient which differs from the ``black-box" approach of \cite{Archer2016} as well as that used by \cite{Titsias2014} and \cite{Kucukelbir2016}. We demonstrate empirically that this estimator has lower variance at the mode and is helpful in improving the stability and precision of the proposed algorithm. This estimator is inspired by \cite{Han2016}, who propose using Gaussian copulas to accommodate models whose posteriors are for instance, skewed, heavy-tailed or multi-modal and hence unsuited to a Gaussian variational approximation. Our idea of introducing sparsity via the Cholesky factor of the precision matrix may prove useful in this context as well. The relationship between the Laplace and the Gaussian variational approximation is discussed in \cite{Opper2009}  while \cite{Challis2013} consider some different parameterizations in terms of the Cholesky. We do not consider Laplace approximations \citep{Rue2009} in this paper since an important advantage of stochastic gradient methods is they are generally amenable to sub-sampling, although this is not always straightforward in complex latent variable models where the local parameters are dependent.

In Section \ref{sec:TL}, we review doubly stochastic variational inference, the approach of \cite{Titsias2014}. Section \ref{sec:prec} describes how the optimization problem can be framed in terms of the precision matrix, develops the algorithm using alternative gradient estimators and discusses the importance of imposing sparsity structure in the precision matrix. The setting of the learning rate in the stochastic gradient algorithm is discussed in Section \ref{sec: learning rate}. In Section \ref{sec:examples}, we illustrate how our approach can be applied to GLMMs and state space models. The performance of our algorithm is investigated using several real data sets. We conclude with a discussion of our major results and findings in Section \ref{sec: conclude}.

\section{Review on doubly stochastic variational inference}\label{sec:TL}
In this section, we provide some general background on variational methods and give a brief review of doubly stochastic variational inference \citep{Titsias2014} as we will be considering a modification of their approach.  

For a Bayesian inference problem, let $\theta$ denote the vector of variables, $p(\theta)$ be the prior and $p(y|\theta)$ the likelihood for observed data $y$. In variational approximation \citep[see, e.g.][]{Bishop2006, Ormerod2010}, an attempt is made to approximate an intractable posterior distribution $p(\theta|y)\propto p(\theta)p(y|\theta)$ using a member of some approximating family. Here we will consider a parametric family with typical element $q_\lambda(\theta)$ where $\lambda$ denotes variational parameters to be chosen. Minimization of the Kullback-Leibler divergence between $q_\lambda(\theta)$ and $p(\theta|y)$ with respect to $\lambda$ can be shown to be equivalent to maximizing a lower bound on the log marginal likelihood $\log  p(y)$ (where $p(y)=\int p(\theta)p(y|\theta)d\theta$), and taking the form 
\begin{equation*}
{\cal L}(\lambda)=\int \log \frac{p(\theta)p(y|\theta)}{q_\lambda(\theta)} q_\lambda(\theta)d\theta.
\end{equation*}
In non-conjugate models, ${\cal L}(\lambda)$ will generally not have a closed form. There has been much recent research concerned with stochastic gradient methods \citep{Robbins1951} able to optimize ${\cal L}(\lambda)$ efficiently in this situation \citep{ji+sw10, nott+tvk12, Paisley2012, Kingma2014, Hoffman2013, Ranganath2014, Rezende2014, Titsias2014, Titsias2015}.  

The method of \citet{Titsias2014} (hereafter TL) is one state of the art method which optimizes ${\cal L}(\lambda)$ using gradient information from the target distribution. Write $h(\theta)=p(\theta)p(y|\theta)$. In the TL method, an approximating distribution of the form 
\begin{equation*}
q(\theta|\mu,L)=|L|^{-1}f(L^{-1}(\theta-\mu))
\end{equation*}
is assumed (so that $\lambda=(\mu,L)$) where $f$ is a fixed density. Here $\mu$ is a vector of parameters of dimension $d$, where $d$ is the dimension of $\theta$, and $L$ is a $d\times d$ lower triangular matrix with positive diagonal elements.  If $f$ is the density of a vector of independent standard normal random variables then $q(\theta|\mu,L)$ is normal, $N(\mu,L L^T)$, and the covariance matrix is being parameterized with $L$ as the Cholesky factor. We will only be considering the case of a multivariate normal approximation in this paper.

The lower bound ${\cal L}(\lambda)={\cal L}(\mu,L)$ is an expectation with respect to $q(\theta|\mu,L)$, but can be written as an expectation with respect to the density $f$. Writing the integral in this way (for the purpose of the stochastic gradient optimization) results in an approach which is able to effectively use gradient information from the target log posterior. More precisely, writing $E_q(\cdot)$ for the expectation with respect to $q(\theta|\mu,L)$ and $E_f(\cdot)$ for the expectation with respect to $f$, we have
\begin{equation} \label{LB1}
\begin{aligned}
{\cal L}(\mu,L) & = E_q\left(\log h(\theta) - \log q(\theta|\mu,L) \right) \\
& = E_f\left(\log h(\mu+Ls) - \log q(\mu+Ls|\mu,L) \right) \\
& = E_f(\log h(\mu+Ls))+\log |L|+K	
\end{aligned}
\end{equation}
where $s = L^{-1}(\theta-\mu)$ is distributed according to the density $f$ and $K$ denotes a term not depending on $\mu,L$. This approach of applying a transformation $\theta = \mu+Ls$ so that the lower bound can be rewritten as an expectation with respect to a fixed density $f$ that does not depend on the variational parameters is sometimes referred to as the ``reparameterization trick" \citep{Kingma2014,Rezende2014, Titsias2014}. The advantage of this approach is that efficient gradient estimators of ${\cal L}(\mu,L)$ can now be constructed by sampling $s$ from $f$ instead of $\theta$ from $q(\theta|\mu,L)$, which has been found to result in estimators with very high variance \citep[see, e.g.][]{Paisley2012}. 

Next, we give expressions for the gradients of ${\cal L}(\mu,L)$ with respect to $\mu$ and $L$. To explain their derivation we need some notation first.  For a scalar valued function $g(x)$ of a vector valued argument $x$, $\nabla_x g(x)$ denotes the gradient vector for the function written as a column vector.  Also, for a scalar valued function $g(A)$ of a matrix $A$, $\nabla_A g(A)$ means $\text{vec}^{-1}(\nabla_{\text{vec}(A)} g(A))$ where, for a $d\times d$ square matrix $A$, $\text{vec}(A)$ is the vector of length $d^2$ obtained by stacking the columns of $A$ underneath each other, and $\text{vec}^{-1}$ is the inverse operation that takes a vector of length $d^2$ and creates a $d\times d$ square matrix by filling up the columns from left to right from the elements of the vector. In addition, we use the following well known result.  For conformably dimensioned matrices $A$, $B$ and $C$, $\text{vec}(ABC)=(C^T \otimes A) \text{vec}(B)$. This implies that we can write $Ls = \text{vec}(ILs) = (s^T \otimes I) \text{vec}(L)$. Then
\begin{equation*}
\nabla_\mu E_f(\log h(\mu+Ls)) = E_f(\nabla_\theta \log h(\mu+Ls)) 
\end{equation*}
and
\begin{equation} \label{exp}
\begin{aligned}
\nabla_{\text{vec}(L)} & E_f(\log h(\mu+Ls)) \\
& = \nabla_{\text{vec}(L)} E_f(\log h(\mu+(s^T\otimes I)\text{vec}(L))) \\
& = E_f((s \otimes I) \nabla_\theta \log h(\mu+Ls)) \\
& = E_f(\text{vec}(\nabla_\theta h(\mu+Ls) s^T)).
\end{aligned}
\end{equation}
The last line of \eqref{exp} just says that 
\begin{equation} \label{gradL}
\nabla_L E_f( \log h(\mu+Ls))=E_f(\nabla_\theta h(\mu+Ls)s^T).
\end{equation} 
Note that entries above the diagonal should be set to zero for the right-hand-side of \eqref{gradL} because $L$ is lower triangular. For the term $\log |L|$ in ${\cal L}(\mu,L)$, we have $\nabla_\mu \log |L|=0$ and $\nabla_L \log | L|=\text{diag}(1/L_{11},\dots, 1/L_{dd})$.  

Once we have expressions for the derivatives of the lower bound as expectations with respect to $f$, we can estimate these gradients unbiasedly using simulations from this distribution (typically based on just a single draw). When the log-likelihood is a sum of a large number of terms, such as in the case of a very large dataset, we can subsample the terms and still construct appropriate unbiased gradient estimates if we desire (hence the name ``doubly stochastic variational inference"). Algorithm \ref{Alg1} shows the basic stochastic gradient method of \citet{Titsias2014}.  The sequence $\rho_t$, $t\geq 1$, in the algorithm is a sequence of learning rates satisfying the Robbins-Monro conditions $\sum_t \rho_t=\infty$, $\sum_t \rho_t^2<\infty$.   
\begin{Algorithm}[htb!]
	\centering
	\parbox{0.38\textwidth}{
		\hrule
		\vspace{1mm}
		Initialize $\mu^{(0)} = 0$ and $T^{(0)} = I_d$.\\
		For $t=1, \dots, N$,
		\begin{enumerate}
			\item Generate $s\sim N(0, I_d)$.
			\item Compute $\theta^{(t)}=\mu^{(t)}+L^{(t)} s$.
			\item Update $\mu^{(t+1)}=\mu^{(t)}+\rho_t \nabla_\theta \log h(\theta^{(t)})$.
			\item Update $L^{(t+1)} = L^{(t)}+\rho_t \{ \nabla_\theta \log h(\theta^{(t)}) s^T$\\
			$+\; \text{diag}(1/L_{11}^{(t)},\dots, 1/L_{dd}^{(t)}) \}$. (Entries above the diagonal are fixed at zero).
		\end{enumerate}
		\vspace{1mm}
		\hrule}
	\caption{Doubly stochastic variational inference algorithm of \citet{Titsias2014}.}\label{Alg1}
\end{Algorithm}

\section{Extension to parametrization of the precision matrix in terms of the Cholesky factor}\label{sec:prec}

When the vector of variables $\theta$ is high-dimensional, allowing $L$ to be a dense matrix is computationally impractical.  An alternative is to assume that $L$ is diagonal, but that loses any ability to capture dependence structure of the posterior. Here we consider an alternative approach where we follow a similar strategy to that of \citet{Titsias2014}, but instead parameterize the inverse covariance (precision) matrix in terms of the Cholesky factor and then impose sparsity on it that reflects conditional independence structure in the model.  

\subsection{Model Specification} \label{sec:model assump}
Consider a model with observations $y= (y_1, \dots, y_n)$, latent variables $b_1, \dots, b_n$ and model parameters $\eta$. Let 
\begin{equation} \label{partition}
\theta = [b_1^T, \dots, b_n^T, \eta^T]^T,
\end{equation}
denote the vector of all variables. We assume that the joint distribution can be written in the form 
\begin{equation} \label{model}
\begin{aligned}
p(y, \theta) & =p(\eta) \left\{ \prod_{i=1}^n p(y_i|b_i, \eta)  \right\} p(b_1, \dots, b_k|\eta)  \\
& \times \left\{ \prod_{i> k}  p(b_i,|b_{i-1}, \dots, b_{i-k},\eta)  \right\}
\end{aligned}
\end{equation}
for some $1 \leq k \leq n$. In this model, $b_i$ is conditionally independent of the other latent variables in the posterior distribution $p(\theta|y)$ given $\eta$ and the neighboring $k$ latent variables.

\subsection{Gaussian variational approximation with sparse precision matrix} \label{GVA}
We consider the variational approximation ($q$) of the posterior to be a multivariate Gaussian distribution $N(\mu, T^{-T}T^{-1})$, where $T$ is a lower triangular matrix with positive diagonal entries. With $f$ being the joint density of a $d$-vector of independent standard Gaussian variables as before, we can write $q(\theta|\mu,T)=|T| f(T^T(\theta-\mu))$. 

Let $\Omega$ denote the precision matrix of the Gaussian distribution. Then $\Omega = TT^T$ and hence $T$ is just the Cholesky factor of $\Omega$. The statistical motivation for imposing sparsity on the Cholesky factor of the precision matrix is as follows. It is well known that for a Gaussian distribution, $\Omega_{ij}=0$ corresponds to variables $i$ and $j$ being conditionally independent given the rest. Also, if $\Omega=TT^T$ where $T$ is lower triangular, proposition 1 of \citet{rothman+lz10} states that if $T$ is row banded then $\Omega$ possesses the same row banded structure. This means that imposing sparsity in $T$ can be useful for reflecting conditional independence relationships in $\Omega$.

For our model in \eqref{model}, let us partition $\Omega$ into blocks $\Omega_{ij}$, $1 \leq i,j \leq n+1$ according to \eqref{partition}. For the Gaussian variational approximation to reflect the conditional independence structure in the posterior, we would like to have $\Omega_{ij} = 0$ for $\{1 \leq i,j \leq n| j < i-k \text{ or } j > i+k\}$, with no constraints on the remaining blocks. Write $T$ for the Cholesky factor partitioned in the same way as $\Omega$ with blocks $T_{ij}$, $1\leq i,j\leq n+1$. Since $T$ is lower triangular, $T_{ij}=0$ if $i<j$ and $T_{ii}$, $1 \leq i \leq n+1$, are lower triangular matrices. If $T_{ij}=0$ for $\{1 \leq j<i \leq n| j < i-k \}$, then $T T^T$ has the sparsity we desire for $\Omega$. The sparsity level of $T$ increases as $k$ decreases. We elaborate later on how the sparse lower triangular structure can be exploited in the generation of $\theta$ from the variational posterior and in gradient computations. This approach is illustrated using generalized linear mixed models and state space models in Section \ref{sec:examples}.

\subsection{Stochastic gradients}
 Similar to the previous case we obtain for the lower bound
\begin{equation} \label{lower_bound}
\begin{aligned}
{\cal L}(\mu,T) & = E_f(\log h(\mu+T^{-T}s) - \log q(\mu+T^{-T}s|\mu,T)) \\
& = E_f(\log h(\mu+T^{-T}s)) -\log |T|+K,
\end{aligned}
\end{equation}
where $s = T^T(\theta-\mu)$ is distributed according to $f$ and $K$ denotes a constant not depending on $\mu,T$. To obtain the gradient of ${\cal L}(\mu,T)$ with respect to $\mu$ and $T$, in addition to the results mentioned in Section \ref{sec:TL}, we need the following result. Denote by $\frac{d {\text{vec}(A^{-1})}}{d \text{vec}(A)}$ the matrix where the $(i,j)$th entry is the partial derivative of the $i$th element of $\text{vec}(A^{-1})$ with respect to the $j$th element of $\text{vec}(A)$. Then 
\begin{equation*}
\frac{d {\text{vec}(A^{-1})}}{d \text{vec}(A)}=-(A^{-T}\otimes A^{-1}).
\end{equation*}
Similar to before, we have
\begin{equation} \label{grad_mu}
\begin{aligned}
\nabla_\mu \mathcal{L} &= \nabla_\mu E_f(\log h(\mu+T^{-T}s)) \\
& = E_f(\nabla_\theta \log h(\mu+T^{-T}s)).
\end{aligned}
\end{equation}
Looking at $T$, 
\begin{equation*}
\begin{aligned}
&\nabla_{\text{vec}(T)} E_f(\log h(\mu+T^{-T}s)) \\
& = E_f\left(\left\{\frac{d\text{vec}(T^{-1})}{d\text{vec}(T)}\right\}^T(I_d \otimes s)  \nabla_\theta \log h(\mu+T^{-T} s)\right) \\
& = -E_f((T^{-1}\otimes T^{-T})(I_d \otimes s) \nabla_\theta \log h(\mu+T^{-T} s)) \\
& = -E_f((T^{-1} \otimes T^{-T}s) \nabla_\theta \log h(\mu+T^{-T}s)) \\
& = -E_f(\text{vec}(T^{-T}s \, (\nabla_\theta \log h(\mu+T^{-T}s) )^T T^{-T})),
\end{aligned}
\end{equation*}
so that 
\begin{equation} \label{T_grad}
\begin{aligned}
\nabla_T \mathcal{L} &= \nabla_T E_f( \log h(\mu+T^{-T}s)) - \nabla_T \log |T|\\
&= - E_f(T^{-T} s (\nabla_\theta \log h(\mu+T^{-T}s))^T T^{-T}) \\
& \quad - \text{diag}(1/T_{11}, \dots, 1/T_{dd}).
\end{aligned}
\end{equation}
Note that for the first term on the right hand side of \eqref{T_grad},  entries above the diagonal should be set to zero because $T$ is a lower triangular matrix.

\subsection{Alternative estimators of the stochastic gradients} \label{sec:alt est}
From \eqref{grad_mu} and \eqref{T_grad}, unbiased estimators of $\nabla_\mu \mathcal{L}$ and $\nabla_T \mathcal{L}$ are given by 
\begin{equation*}
\begin{aligned}
\hat{g}_{\mu,1} &= \nabla_\theta \log h(\mu+T^{-T}s) \quad \text{and} \nonumber \\ 
\hat{g}_{T,1} &= -T^{-T} s (\nabla_\theta \log h(\mu+T^{-T}s))^T T^{-T}) \\
& \quad - \text{diag}(1/T_{11}, \dots, 1/T_{dd})
\end{aligned}
\end{equation*}
respectively, where $s$ is generated from $N(0, I_d)$. In deriving these estimators, we have evaluated the term $E_f(\log q(\mu+T^{-T}s|\mu,T))$ in the lower bound analytically. However, alternative estimators can be derived by approximating this term instead of using its analytical form. As 
\begin{equation*}
\nabla_\theta \log q(\theta) = -TT^T (\theta - \mu) = -Ts,
\end{equation*}
we have from \eqref{lower_bound}, 
\begin{equation*}
\begin{aligned}
\nabla_\mu \mathcal{L} &= \nabla_\mu E_f(\log h(\mu+T^{-T}s) -\log q(\mu+T^{-T}s))  \\
& = E_f(\nabla_\theta \log h(\mu+T^{-T}s) -\nabla_\theta \log q(\mu+T^{-T}s)) \\
& = E_f(\nabla_\theta \log h(\mu+T^{-T}s) + Ts).
\end{aligned}
\end{equation*}
Similarly, 
\begin{equation*}
\begin{aligned}
\nabla_T \mathcal{L} &= \nabla_TE_f( \log h(\mu+T^{-T}s) -\log q(\mu+T^{-T}s) )\\
&= - E_f(T^{-T} s (\nabla_\theta \log h(\mu+T^{-T}s)  \\
& \quad -\nabla_\theta \log q(\mu+T^{-T}s))^T T^{-T}) \\
& =  - E_f(T^{-T} s (\nabla_\theta \log h(\mu+T^{-T}s) + Ts)^T T^{-T}),
\end{aligned}
\end{equation*}
Thus alternative unbiased estimators of $\nabla_\mu \mathcal{L}$  and $\nabla_T \mathcal{L}$ are given by 
\begin{equation*}
\begin{aligned}
\hat{g}_{\mu,2} &= \nabla_\theta \log h(\mu+T^{-T}s) + Ts  \quad \text{and} \\
\hat{g}_{T,2} &=- T^{-T} s (\nabla_\theta \log h(\mu+T^{-T}s) + Ts)^T T^{-T}.
\end{aligned}
\end{equation*}

\begin{figure*}[htb!]
	\centering
	\includegraphics[width=\textwidth]{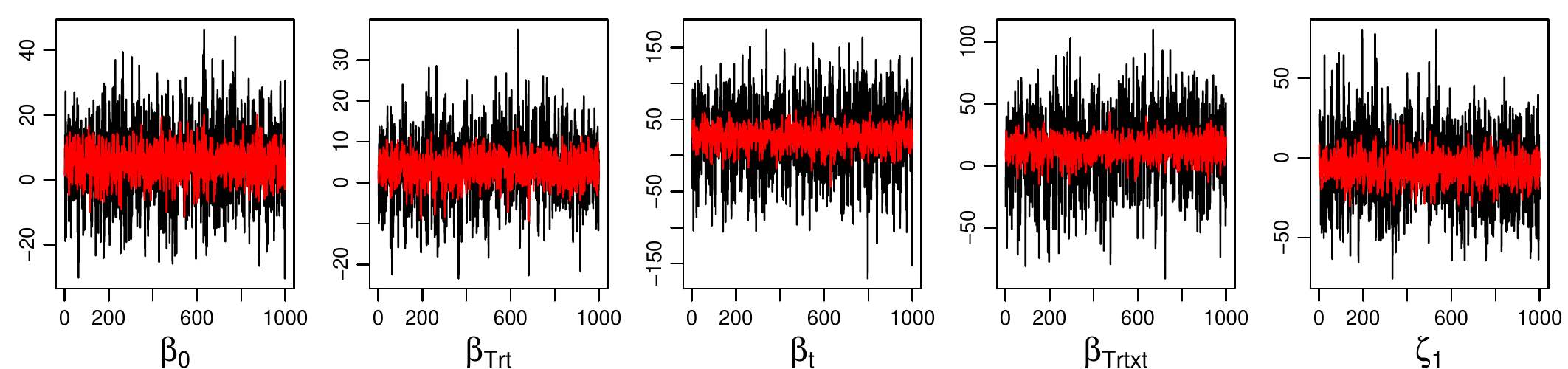}
	\caption{Toenail data: estimates of $\nabla_\mu \mathcal{L}$ given by $\hat{g}_{\mu,1}$ (black) and $\hat{g}_{\mu,2}$ (red) at the mode for a subset of the variables. \label{grad_compare}}
\end{figure*}

In our experiments, we observe that the estimators $\hat{g}_{\mu,2}$ and $\hat{g}_{T,2}$ seem to provide approximations with lower variance (\cite{Salimans2013} and \cite{Han2016} note related phenomena). As an example, for the toenail dataset in Section \ref{toenail}, we compare in Figure \ref{grad_compare} estimates of $\nabla_\mu \mathcal{L}$ given by $\hat{g}_{\mu,1}$ (black) and $\hat{g}_{\mu,2}$ (red) for a subset of the variables. This is done by fixing $\mu$ and $T$ at the mode and computing the gradient estimates of $\mu$ at 1000 random variates $s \sim N(0, I_d)$. Figure \ref{grad_compare} shows clearly that there is much greater variation in the estimates computed using $\hat{g}_{\mu,1}$  as compared to $\hat{g}_{\mu,2}$. This suggests that using the alternative estimators $\hat{g}_{\mu,2}$ and $\hat{g}_{T,2}$  will result in a more stable algorithm with better convergence and greater precision.

Below, we provide some intuition for this observation. Suppose the density that we are approximating is close to a Gaussian distribution with mean $\mu^*$ and precision $T^*{T^*}^T$, that is, $ h(\theta) \approx N(\mu^*, {T^*}^{-T}{T^*}^{-1})$. Then $\nabla_\theta \log h(\theta) \approx -T^*{T^*}^T (T^{-T}s + \mu - \mu^*)$. When we are close to the mode, $T\approx T^*$, $\mu \approx \mu^*$ and
\begin{equation*}
\begin{aligned}
\hat{g}_{\mu,1} &\approx -T^*s, \\
\hat{g}_{T,1} & \approx {T^*}^{-T} ss^T - \text{diag}(1/T^*_{11}, \dots, 1/T^*_{dd}),
\end{aligned}
\end{equation*}
while
\begin{equation*}
\hat{g}_{\mu,2} \approx 0, \quad  \hat{g}_{T,2} \approx 0.
\end{equation*}
Thus, for $\hat{g}_{\mu,2}$, the contributions to the gradients from $\nabla_\theta \log h(\theta)$ and $\nabla_\theta \log q(\theta)$ cancel out. As  $\hat{g}_{\mu,2}$ is a factor of $\hat{g}_{T,2}$, $\hat{g}_{T,2} \approx 0$ when $\hat{g}_{\mu,2} \approx 0$. However, $\hat{g}_{\mu,1}$ and $\hat{g}_{T,1}$ are still subjected to the randomness in $s$ around the mode. Thus we prefer to use the estimators $\hat{g}_{\mu,2}$ and $\hat{g}_{T,2}$, which do not incur any additional computation except for the term $Ts$.

\subsection{Uniqueness of the Cholesky factor}

We note that in Algorithm 1, the updates of $L$ do not ensure that the diagonal entries are positive. While this does not result in any computational issues, we prefer to add in the following step to ensure that the diagonal entries of $T$ are positive. This helps to ensure uniqueness of the Cholesky factor and reduces the possibility of multiple local modes which is an important issue especially in the high-dimensional problems considered here. To achieve this aim. We introduce the lower triangular matrix $T'$ where $T'_{ij}$ = $T_{ij}$ if $i \neq j$ and $T'_{ii} =\log(T_{ii})$. We compute the stochastic gradient updates for $T'$ whose entries are unconstrained. The gradient $\hat{g}_{T',2} = \hat{g}_{T,2}$ for all non-diagonal entries. Diagonal entries of $\hat{g}_{T',2} $ can be computed by multiplying the diagonal entries of $\hat{g}_{T,2} $ by the diagonal entries of $T$.

The modification of the doubly stochastic variational inference algorithm, in terms of the Cholesky factor of the precision matrix, is summarized in Algorithm \ref{Alg2}.
\begin{Algorithm}[htb!]
\centering
\parbox{0.38\textwidth}{
\hrule
\vspace{1mm}
Initialize $\mu^{(0)} = 0$, $T^{(0)} = I_d$ and $T'^{(0)} = 0$.\\
For $t=1, \dots, N$,
\begin{enumerate}
\item Generate $s\sim N(0, I_d)$.
\item Compute $\theta^{(t)}=\mu^{(t-1)}+{T^{(t-1)}}^{-T} s$.
\item Compute $g_\mu^{(t)} = \nabla_\theta \log h(\theta^{(t)}) + T^{(t-1)}s$.
\item Update $\mu^{(t)}=\mu^{(t-1)}+\rho_t \; g_\mu^{(t)} $.
\item Set $g_{T'}^{(t)} = -{T^{(t-1)}}^{-T} s \; ( {T^{(t-1)}}^{-1} g_\mu^{(t)})^{T}$.
\item Set $\text{diag}(g_{T'}^{(t)}) =  \text{diag}(g_{T'}^{(t)}) \odot \text{diag}(T^{(t-1)})$.
\item Update $T'^{(t)}=T'^{(t)} + \rho_t \; g_{T'}^{(t)} $.
\item Update $T^{(t)} = T'^{(t)}$.
\item Update $\text{diag}(T^{(t)}) = \exp(\text{diag}(T'^{(t)}))$.
\end{enumerate}
\hrule}
\caption{Modified doubly stochastic variational inference algorithm parameterized in terms of the Cholesky factor of the precision matrix.}\label{Alg2}
\end{Algorithm}

Now let us consider sparsity in the Cholesky factor $T$. Suppose some elements of $T$ are fixed at zero. Then Algorithm 2 remains the same, except that only the subset of elements of $T$ which are not fixed at zero are stored and updated. Note that in step 2, if $T$ is a sparse matrix, we can compute $T^{-T}s$ by solving the linear system $T^T x=s$ for $x$. This can be done very efficiently because $T^T$ is upper triangular and sparse. Similarly, in computing the update at step 5, we need to compute the vector $T^{-1}g_\mu$. This can also be computed by solving the linear system $T x=g_\mu$, which is again easy because $T$ is a sparse lower triangular matrix. So even in very high dimensions, if $T$ is appropriately sparse, Algorithm \ref{Alg2} can be implemented in a way that is efficient in terms of both memory storage requirements and CPU time.

\section{Setting the learning rate and stopping criterion in the stochastic optimization} \label{sec: learning rate}
\subsection{Learning rate}
The setting of appropriate learning rates in stochastic gradient algorithms is a highly challenging problem. The choice of learning rate determines not only the rate of convergence but also the quality of the optimum attained. Learning rates that are too high causes the algorithm to diverge while rates that are too low results in slow learning and can lead to ``apparent convergence", a situation where parameters appear to have converged due to diminishing step-size \citep[see, e.g.][]{Powell2011}. \cite{Spall2003} suggests a step-size sequence which satisfies the theoretical conditions for convergence \citep{Robbins1951}. This takes the form $A_1/(t + A_2)^{A_3}$ where $t$ denotes the iteration, and $A_1 \geq 1$, $A_2\geq 0$ and $0.5 < A_3 \leq 1$ are constants to be tuned. However, we find that it is difficult to hand-tune this learning rate for use in Algorithm 2, as the problems considered are high-dimensional in nature and the parameters $\{\mu, T\}$ converge at different rates. For instance, \cite{Titsias2014} scaled down the learning rate of $\mu$ by 0.1 when using for $L$ in Algorithm 1. It is also likely that the parameters have different ``scale", especially when some of the constrained parameters have to be transformed to the real line. These considerations increase the need for learning rates that are adaptive and parameter-specific. Several adaptive step-size sequences (e.g. \cite{Duchi2011}, \cite{Ranganath2013}, \citealp{Kucukelbir2016}) have been proposed. We find that the ADADELTA method \citep{Zeiler2012}, in particular, worked very well with Algorithm 2 and we use it for all the examples. For consistency, we also used ADADELTA to compute the step-size for Algorithm 1. While ADADELTA has worked well in our experiments, we have only worked on a limited number of datasets and it is likely that other learning rates may yield better performance. From our observations, the performance of learning rates tend to be problem-dependent.

The ADADELTA method takes into consideration the scale of the parameters by incorporating second order information through a Hessian approximation.  Suppose at iteration $t$, a parameter $x$ is updated as $x^{(t)} = x^{(t-1)} + \Delta x^{(t)}$, where $\Delta x^{(t)} = \omega g_x^{(t)}$ and $g_x^{(t)}$ is the gradient. The step-size $\omega$ is computed as 
\begin{equation} \label{optimalstepsize}
\omega = \frac{\sqrt{E[\Delta_x^2]^{(t-1)} + \epsilon}}{\sqrt{E[g_x^2]^{(t)} + \epsilon}},
\end{equation}
where $E[\Delta_x^2]^{(t-1)}$ and $E[g_x^2]^{(t)}$ are exponentially decaying averages of ${\Delta x^{(t)}}^2$ and ${g_x^{(t)}}^2$, and $\epsilon$ is a small positive constant added to ensure the denominator is positive and the initial step-size is nonzero. The terms $E[\Delta_x^2]^{(t)}$ and $E[g_x^2]^{(t)}$ are updated as
\begin{equation*}
\begin{aligned}
E[\Delta_x^2]^{(t)} &=\rho E[\Delta_x^2]^{(t-1)} + (1-\rho) {\Delta_x^{(t)}}^2  \;\;\text{and}\;\;  \\
E[g_x^2]^{(t)} &=\rho E[g_x^2]^{(t-1)} + (1-\rho) {g_x^{(t)}}^2
\end{aligned}
\end{equation*}
at each iteration where $\rho$ is a decaying constant. The motivation of this approach comes from Newton-Raphson algorithms where it is well known that the inverse of the Hessian matrix provides an optimal or near-optimal step-size sequence \cite[see, e.g.][]{Spall2003}. ADADELTA approximates the Hessian by taking $\frac{1}{f''(x)} \approx \frac{\Delta x}{f'(x)}$, hence the form of $\omega$ in \eqref{optimalstepsize}. To apply ADADELTA, we modify Algorithm 2 as outlined below. 
\begin{Algorithm}[h]
\caption*{Algorithm 2 with ADADELTA.}
\centering
\parbox{0.49\textwidth}{
\hrule
\vspace{1mm}
Initialize $\mu^{(0)}=0$, $T^{(0)}=I_d$, $T'^{(0)}=0$, \\ [1mm]
$E[g_\mu^2]^{(0)} = E[\Delta_\mu^2]^{(0)} = 0$, $E[g_{T'}^2]^{(0)} =E[\Delta_{T'}^2]^{(0)} = 0$.\\ [1mm]
For $t=1, \dots, N$,
\begin{enumerate}[itemsep=3pt]
\item Generate $s\sim N(0,I_d)$.
\item $\theta^{(t)}=\mu^{(t-1)}+{T^{(t-1)}}^{-T} s$.
\item $g_\mu^{(t)} = \nabla_\theta \log h(\theta^{(t)})+ T^{(t-1)}s$ .
\item Accumulate gradient $E[g_\mu^2]^{(t)} =\rho E[g_\mu^2]^{(t-1)} + (1-\rho) {g_\mu^{(t)}}^2$.
\item Compute change $\Delta_\mu^{(t)} = \frac{\sqrt{E[\Delta_\mu^2]^{(t-1)} + \epsilon}}{\sqrt{E[g_\mu^2]^{(t)} + \epsilon}} g_\mu^{(t)}$
\item Accumulate change \\
$E[\Delta_\mu^2]^{(t)} =\rho E[\Delta_\mu^2]^{(t-1)} + (1-\rho){ \Delta_\mu^{(t)}}^2$.
\item $\mu^{(t)}=\mu^{(t-1)}+ \Delta_\mu^{(t)}$
\item $g_{T'}^{(t)} = - {T^{(t-1)}}^{-T} s  ( {T^{(t-1)}}^{-1}g_\mu^{(t)} )^T $.
\item Set $\text{diag}(g_{T'}^{(t)}) =  \text{diag}(g_{T'}^{(t)}) \odot \text{diag}(T^{(t-1)})$.
\item Accumulate gradient \\
$E[g_{T'}^2]^{(t)} =\rho E[g_{T'}^2]^{(t-1)} + (1-\rho) {g_{T'}^{(t)}}^2$.
\item Compute change $\Delta_{T'}^{(t)} = \frac{\sqrt{E[\Delta_{T'}^2]^{(t-1)} + \epsilon}}{\sqrt{E[g_{T'}^2]^{(t)} + \epsilon}} g_{T'}^{(t)}$
\item Accumulate change \\
$E[\Delta_{T'}^2]^{(t)} =\rho E[\Delta_{T'}^2]^{(t-1)} + (1-\rho){ \Delta_T^{(t)}}^2$.
\item $T'^{(t)}=T'^{(t-1)} + \Delta_{T'}^{(t)}$.
\item Update $T^{(t)} = T'^{(t)}$.
\item Update $\text{diag}(T^{(t)}) = \exp(\text{diag}(T'^{(t)}))$.
\end{enumerate}
\hrule}
\end{Algorithm} 
Note that the step-size for $\mu$ and $T$ are different. As $T$ is a sparse matrix, we find that it is more efficient to perform steps 8--12 in vector-form and to store only the non-zero elements of $g_T^{(t)}$, $\Delta_T^{(t)}$, $E[g_{T'}^2]^{(t)}$, $E[\Delta_{T'}^2]^{(t)}$. We let $\epsilon = 10^{-6}$ and $\rho=0.95$, the setting recommended by \cite{Zeiler2012}. We note that Algorithm 2 is more sensitive to $\rho$ when the estimators $\hat{g}_{\mu,1}$ and $\hat{g}_{T,1}$ are used as compared to the alternative estimators $\hat{g}_{\mu,2}$ and $\hat{g}_{T,2}$.

\subsection{Stopping Criterion} \label{sec: stop crit}
In variational algorithms, the lower bound is commonly used as an objective function to check for convergence. When the updates are deterministic, the lower bound is guaranteed to increase after each cycle and the algorithm can be terminated when the increase in the lower bound is negligible. In Algorithms 1 and 2, the updates are stochastic and so the lower bound is not guaranteed to increase at each iteration. Computing the lower bounds in \eqref{LB1} and \eqref{lower_bound} also requires evaluating the expectations with respect to the variational approximation $q$. It is straightforward, however, to obtain an unbiased estimate of the lower bound at each iteration. Let  $s$ be a random variate generated from $N(0, I_d)$. From \eqref{LB1}, an unbiased estimate of the lower bound for Algorithm 1 is given by
\begin{equation*}
\begin{aligned}
\hat{\mathcal{L}} &= \log h(\mu+Ls) - \log q(\mu+Ls|\mu, L) \\
& = \log h(\mu+Ls) +\frac{d}{2}\log (2\pi) + \log|L| + \frac{1}{2} s^Ts
\end{aligned}
\end{equation*}
Similarly, an unbiased estimate of the lower bound for Algorithm 2 is given by
\begin{equation*}
\begin{aligned}
\hat{\mathcal{L}} %&= \log h(\mu+T^{-T}s) - \log q(\mu+T^{-T}s|\mu,T) \\
& = \log h(\mu+T^{-T}s) +\frac{d}{2}\log (2\pi) - \log|T| + \frac{1}{2} s^Ts.
\end{aligned}
\end{equation*}
Here we do not evaluate the expectation of the last term $\frac{1}{2}s^Ts$ analytically so that the randomness in $s$ will cancel out between the first and the last term when we are close to the mode (see similar argument is given in Section \ref{sec:alt est}). 

As the estimate $\hat{L}$ is stochastic, we consider instead the average of $\hat{L}$ over the past $F$ iterations, say $\bar{\mathcal{L}}$, to minimize variability. We compute $\bar{\mathcal{L}}$ after every $F$ iterations and keep a record of the maximum value of $\bar{\mathcal{L}}$ attained thus far, say $\bar{\mathcal{L}}_{\text{max}}$. The algorithm is terminated when $\bar{\mathcal{L}}$ falls below $\bar{\mathcal{L}}_{\text{max}}$ more than $M$ consecutive times. This may imply either that the algorithm has converged and hence the lower bound estimates are just bouncing around or the algorithm is diverging and the estimates of the lower bound are deteriorating. We say that the algorithm is ``diverging" if $\bar{\mathcal{L}}$ is tending towards $-\infty$. In Section \ref{sec:examples}, we adopt rather conservative values of $F=2500$ and $M=3$ to avoid the dangers of stopping prematurely \citep[see, e.g.][]{Booth1999}. Alternative stopping criteria can also be constructed by examining the relative change in the parameter updates from successive iterations or the magnitude of the gradients of the parameters \citep[see, e.g.][]{Spall2003}.

\section{Applications}\label{sec:examples}
In this section, we illustrate how we can impose sparsity in $T$ via Algorithm 2 using appropriate posterior conditional independence relationships for generalized linear mixed models (GLMMs) and state space models (SSMs). We code Algorithms 1 and 2 in Julia Version 0.5.0 (\url{http://julialang.org/}) and make use of its functions for sparse matrix representations to store and perform operations on high-dimensional sparse matrices efficiently. 

\begin{table*}[htb!]
\centering \ra{1.1}
\begin{tabular}{@{}lcccccc@{}}
\hline
& Size  & \multicolumn{2}{c}{ADVI (Stan)} & \multicolumn{2}{c}{Algorithm 1} & Algorithm 2 \\ 
& ($n$) & mean-field & unrestricted & mean-field & unrestricted & \\ \hline
Epilepsy (Model I) &  59 & 2  & 1 & 7 (350) & 18 (400) & 20 (725) \\ 
Epilepsy (Model II) &  59  & 4 & 4 & 9 (350) & 45 (550) & 32 (850) \\
Toenail  & 294 & 4 & 15 & 18 (150) & 135 (325) & 88 (650)  \\
Polypharmacy & 500 & 7 & 74 & 30 (150) & 262 (250) & 56 (225) \\
GBP/USD exchange rate & 945 & 10 & diverge & 10 (600) & diverge & 43 (750) \\ 
DEM/USD exchange rate & 1866 & 18 & diverge & 23 (700) & diverge & 77 (700) \\ \hline
\end{tabular}
\caption{Runtime (in seconds) of ADVI (Stan) and Algorithms 1 and 2 (Julia) for datasets of different sizes. The number of iterations (in hundreds) used in Algorithms 1 and 2 are given in brackets.}
\label{runtimes}
\end{table*}

We compare the variational approximations with posteriors obtained through long runs of MCMC (regarded as ground-truth). In all examples, fitting via MCMC was performed in RStan (\url{http://mc-stan.org/interfaces/rstan}) and the same priors are used in MCMC and variational approximations. For MCMC, we use 50,000 iterations in each example and the first half is discarded as burn-in. A thinning factor of 5 was applied and the remaining 5000 samples are used to estimate the posterior density.

We note that Algorithm 1 can also be readily implemented in Stan using automatic differentiation variational inference \citep[ADVI,][]{Kucukelbir2016}. Hence we have also included the results from ADVI for comparison. However, there are some differences between our implementation of Algorithm 1 in Julia and that in Stan, namely, the learning rate and stopping criterion are different and we impose the additional restriction that the diagonal elements in $L$ must be positive.

Table \ref{runtimes} shows the runtimes for ADVI and Algorithms 1  and 2 for the datasets considered in this section. We use the terms ``mean-field" to refer to the case where $L$ is a diagonal matrix and ``unrestricted" when $L$ is a full lower triangular matrix. All experiments are run on a Intel Core i5 CPU@ 3.20GHz 8.0GB Ram.

\subsection{Generalized linear mixed models} \label{sec:GLMM}
Here we consider GLMMs where $y_i=(y_{i1},\dots,y_{in_i})^T$ is the set of responses for the $i$th subject, $X_{ij}$ and $Z_{ij}$ are vectors of predictors for $y_{ij}$, $\mu_{ij}=E(y_{ij})$, and $g(\cdot)$ is a smooth invertible link function. Let 
\begin{gather*}
g (\mu_{ij}) = X_{ij}^T \beta + Z_{ij}^T b_i \;\;\text{for}\;\; i=1, \dots, n, \;\;j=1, \dots, n_i , \\
b_i \sim N(0, G) \;\;\text{for}\;\; i=1, \dots, n, \\
\beta \sim N(0, \sigma_\beta^2I_k),  
\end{gather*}
where $\beta$ is a vector of fixed effects parameters and $b_i$ is a random effect for the $i$th subject. Here we consider binary responses, where $y_{ij} \sim \text{Bernoulli}(\mu_{ij})$ with the logit link function $g (\mu_{ij}) = \log \frac{\mu_{ij}}{1-\mu_{ij}}$, and count responses, where $y_{ij} \sim \text{Poisson}(\mu_{ij})$ with the log link function $g (\mu_{ij}) = \log (\mu_{ij})$. Variational methods have been considered for efficient computation in GLMMs by \cite{Ormerod2012, Tan2013, Tan2014, Lee2016b, Lee2016a}, among others.

We parameterize the elements of the random effects covariance matrix $G$ so that they are unconstrained and so that a normal variational posterior approximation is reasonable. Let $G=WW^T$, where $W$, the Cholesky factor of $G$, is a $p \times p$ lower triangular matrix with positive diagonal entries. Let $W^*$ denote the matrix for which $W_{ii}^* = \log(W_{ii})$ and $W_{ij}^* = W_{ij}$ if $i \neq j$. Write $\zeta = \text{vech}(W^*)$, where $\text{vech}$ is the operation that transforms the lower triangular part of a square matrix into a vector by stacking elements below the diagonal column by column. We assume a normal prior, $\zeta \sim N(0, \sigma_\zeta^2I_{p(p+1)/2})$. 

The vector of variables in the model is given by $\theta$ as defined in \eqref{partition}, where $\eta = (\beta^T, \zeta^T)^T$ and the length of $\theta$ is $d = pn+ k+p(p+1)/2$. The joint distribution can be written as 
\begin{equation*}
p(y, \theta) = h(\theta) = p(\beta) p(\zeta) \prod_{i=1}^n  \left\{ p(b_i|\zeta) \prod_{j=1}^{n_i} p(y_{ij}|\beta, b_i)  \right\}.
\end{equation*}
For this model, note that $b_i$ and $b_j$ are conditionally independent in the posterior distribution for $i\neq j$ given $\eta$. For the GLMM, the sparsity structure imposed on $T$ and hence $\Omega$ is illustrated in \eqref{Tstructure_GLMM}. Our Algorithm 2 can efficiently learn a $T$ with such a structure.
\begin{equation} \label{Tstructure_GLMM}
\begin{aligned}
T &= \begin{bmatrix}
T_{11} & 0 & \ldots & 0 & 0 \\
0 & T_{22} & \ldots & 0 & 0 \\
\vdots & \vdots & \ddots & \vdots & \vdots \\
0 &  0 & \ldots & T_{nn} & 0 \\
T_{n+1,1} & T_{n+1,2} & \ldots & T_{n+1,n} & T_{n+1,n+1} \\
\end{bmatrix}, \\
\Omega &= \begin{bmatrix}
\Omega_{11} & 0 & \ldots & 0 & \Omega_{1,n+1} \\
0 & \Omega_{22} & \ldots & 0 & \Omega_{2,n+1} \\
\vdots & \vdots & \ddots & \vdots & \vdots \\
0 &  0 & \ldots & \Omega_{nn} & \Omega_{n,n+1} \\
\Omega_{n+1,1} & \Omega_{n+1,2} & \ldots & \Omega_{n+1,n} & \Omega_{n+1,n+1} \\
\end{bmatrix}
\end{aligned}
\end{equation}
For the GLMM, using a full rank lower triangular matrix $L$ in Algorithm 1 requires updates of $\mathcal{O}(n^2p^2)$ elements at each iteration while Algorithm 2 only requires $\mathcal{O}(np^2)$ updates (assuming $k$ and $p$ are small). Hence the efficiency of Algorithm 2 as compared to Algorithm 1 (unrestricted) increases rapidly with the number of subjects in the dataset as can be seen from Table \ref{runtimes}. There is only a slight computational overhead in using Algorithm 2 as compared to a diagonal matrix $L$ in Algorithm 1, which requires $\mathcal{O}(np)$ updates. However, Algorithm 2 reflects the posterior dependency structure and hence has the potential to provide better estimates. Next, we investigate the performance of Algorithm 2 on several data sets. We set $\sigma_\beta^2 = \sigma_\zeta^2 =100$ throughout. The gradient of $\log h(\theta)$ is derived in \ref{GLMM_deriv}.

\subsubsection{Epilepsy data}
The epilepsy data of \cite{Thall1990} includes $n=59$ epileptics who were randomized to a new drug, progabide (Trt=1) or a placebo (Trt=0) in a clinical trial. 
The response is given by the number of seizures patients have during four follow-up periods. Other covariates include the logarithm of age (Age), the logarithm of $\frac{1}{4}$ the number of baseline seizures (Base), Visit (coded as $\text{Visit}_1=-0.3$, $\text{Visit}_2=-0.1$, $\text{Visit}_3=0.1$ and $\text{Visit}_4=0.3$), and a binary variable V4 which is 1 for the fourth visit and 0 otherwise. We center the covariate Age and replace $\text{Age}_i$ by $\text{Age}_i-\text{mean(Age)}$. Consider the following two models from \cite{Breslow1993}. Model I is a Poisson random intercept model where
\begin{equation*}
\begin{aligned}
\log \mu_{ij} &= \beta_0+\beta_{\text{Base}} \text{Base}_i+\beta_{\text{Trt}} \text{Trt}_i +\beta_{\text{Age}} \text{Age}_i  \\
& \quad + \beta_{\text{Base} \times \text{Trt}} \text{Base}_i \times, \text{Trt}_i+\beta_{\text{V4}} \text{V4}_{ij}+b_i
\end{aligned}
\end{equation*}
for $i=1,...,n$, $j=1,...,4$. Model II is a Poisson random intercept and slope model where
\begin{equation*}
\begin{aligned}
\log \mu_{ij} &= \beta_0+\beta_{\text{Base}} \text{Base}_i+\beta_{\text{Trt}} \text{Trt}_i +\beta_{\text{Age}} \text{Age}_i\\
& \quad + \beta_{\text{Base} \times \text{Trt}} \text{Base}_i \times \text{Trt}_i +\beta_{\text{Visit}} \text{Visit}_{ij}  \\
& \quad +b_{i1} + b_{i2} \text{Visit}_{ij}
\end{aligned}
\end{equation*}
for $i=1,...,n$, $j=1,...,4$.

\begin{figure*}[htb!]
\centering
\includegraphics[width=\textwidth]{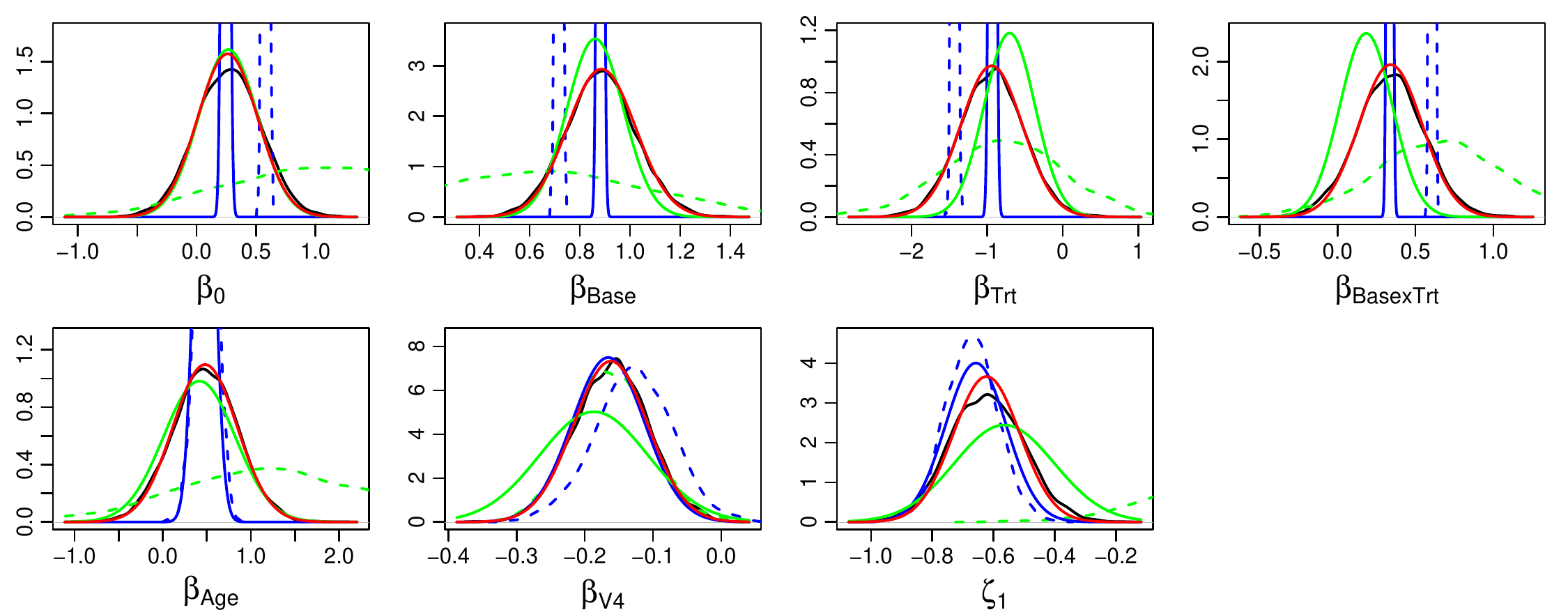}
\includegraphics[width=0.9\textwidth]{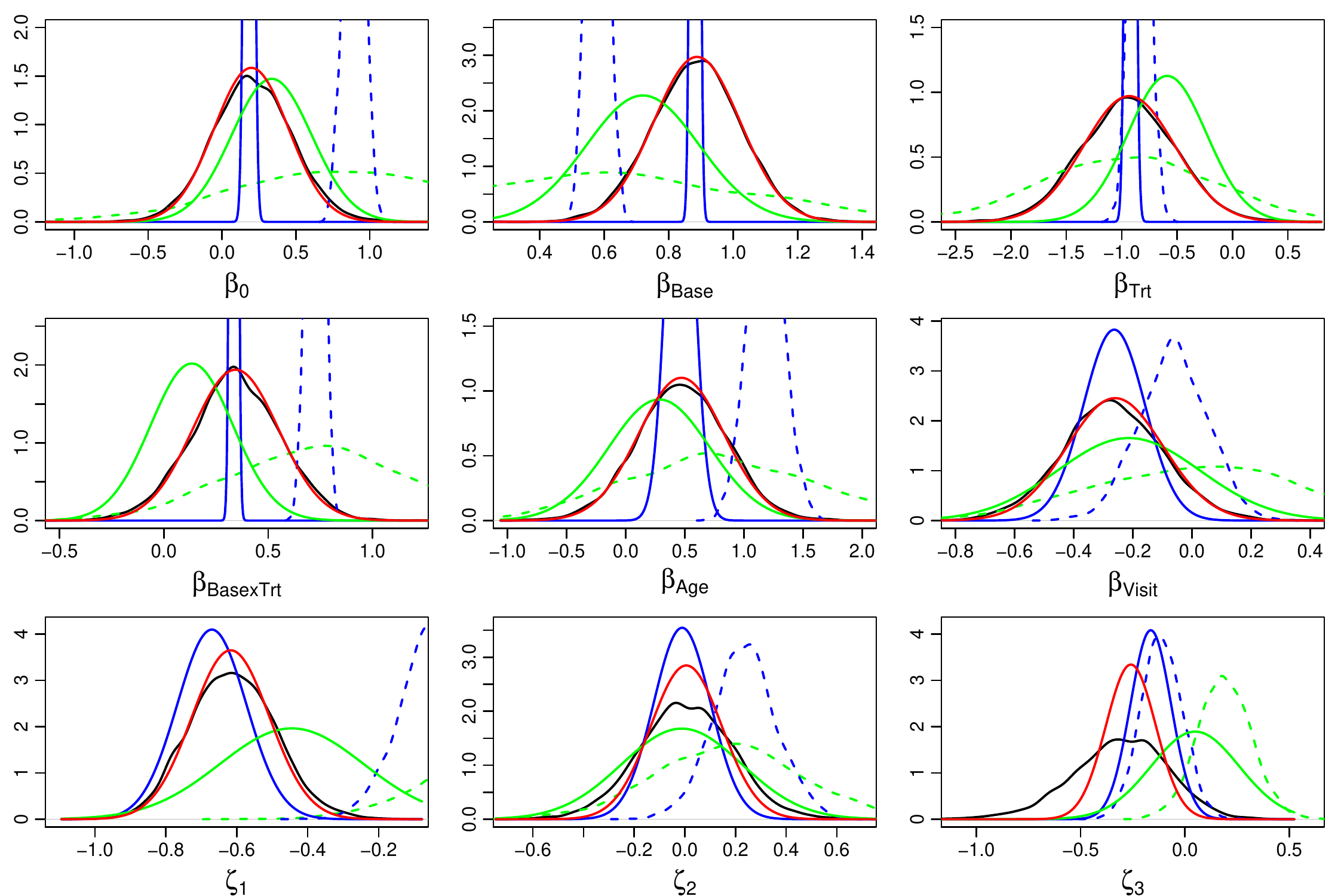}
\caption{Epilepsy data: posterior distributions of $\beta$ and $\zeta$ estimated using ADVI (dotted, blue for mean-field and green for unrestricted), Algorithm 1 (blue for mean-field and green for unrestricted), Algorithm 2 (red) and MCMC (black) for Model I (first two rows) and Model II (last two rows). \label{epilepsy_posteriors}}
\end{figure*}

We apply ADVI and Algorithms 1 and 2 on these two models. Runtimes are given in Table \ref{runtimes} and the estimated marginal posteriors of $\beta$ and $\zeta$ are shown in Figure \ref{epilepsy_posteriors}. Algorithm 1 (mean-field) converged quickly for both models while the runtime of Algorithm 1 (unrestricted) doubled with the inclusion of a second random effect. For this dataset, Algorithm 2 performed better than the mean-field and unrestricted approximations. It produces very good approximations of the marginal posteriors of $\beta$, but is overconfident in estimating the marginal posteriors of $\zeta$ in Model II. The variational posteriors from Algorithm 1 (mean-field) are accurate in the mean but the variance is underestimated, quite severely in some cases. 

\begin{figure*}[htb!]
\centering
\includegraphics[width=\textwidth]{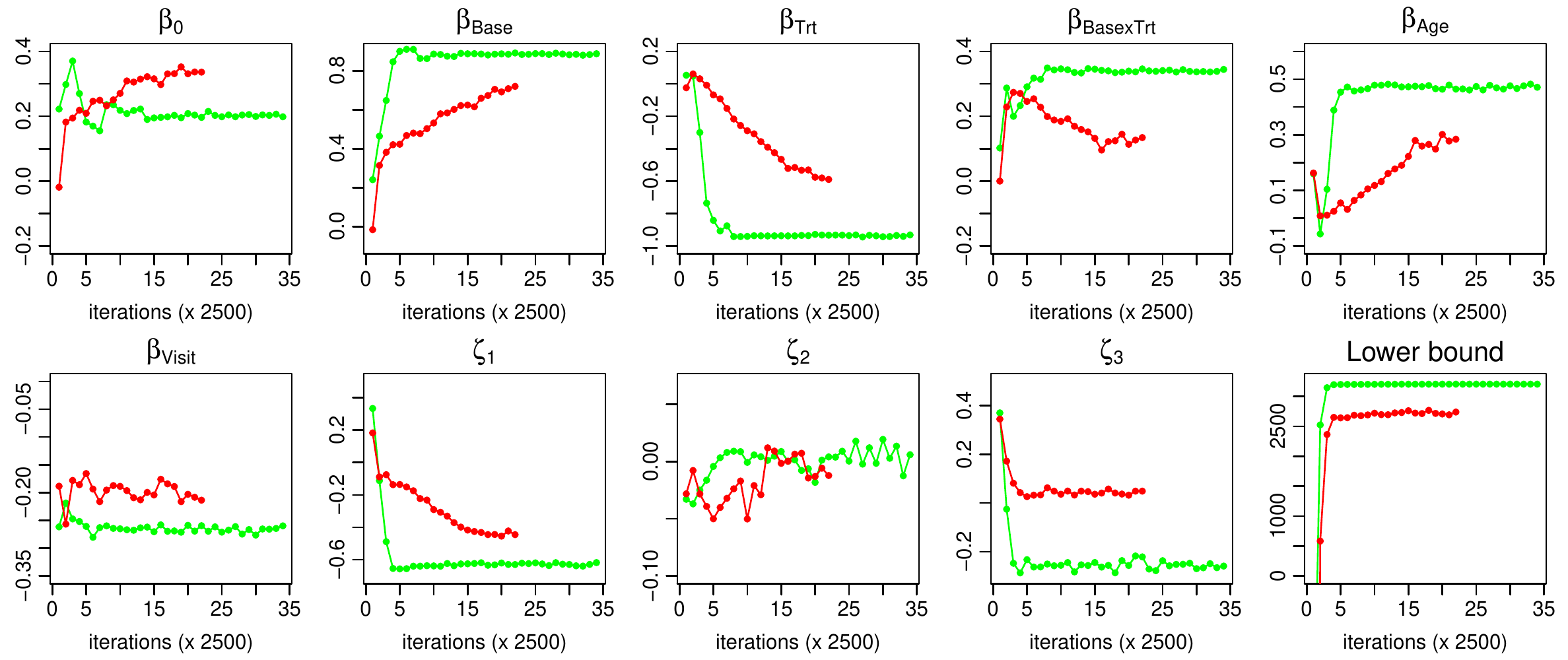}
\caption{Epilepsy data: Mean ($\mu$) iterates corresponding to $\beta$ and $\zeta$ and the averaged lower bound ($\bar{\mathcal{L}}$) from Algorithm 1 with unrestricted lower triangular matrix $L$ (green) and Algorithm 2 (red) for Model II. \label{epilepsy_convergence}}
\end{figure*}

Figure \ref{epilepsy_convergence} shows the iterates of the mean parameter $\mu$ corresponding to $\beta$ and $\zeta$ and the averaged lower bound ($\bar{\mathcal{L}}$) for Model II. For Algorithm 1 (unrestricted), it appears that some of the parameters have yet to stabilize even though the lower bound has reached stationarity. 

\subsubsection{Toenail data} \label{toenail}
This dataset compares two oral anti-fungal treatments for toenail infection \citep{DeBacker1998} and contains information for 294 patients who are evaluated at seven visits. Some patients did not attend all planned visits and there were a total of 1908 measurements. The patients were randomized into two treatment groups, one receiving 250 mg of terbinafine per day (Trt=1) and the other 200 mg of itraconazole per day (Trt=0). The time in months ($t$) that they arrived for visits was recorded and the binary response variable (onycholysis) indicates the degree of separation of the nail plate from the nail-bed (0 if none or mild, 1 if moderate or severe). We consider the logistic random intercept model,
\begin{equation*}
\text{logit}(\mu_{ij}) = \beta_0+\beta_{\text{Trt}} \text{Trt}_i+\beta_t t_{ij}
+\beta_{\text{Trt} \times t} \text{Trt}_i \times t_{ij}+u_i,
\end{equation*}
for $i=1,...,294$, $1\leq j \leq 7$. 

\begin{figure*}[htb!]
\centering
\includegraphics[width=\textwidth]{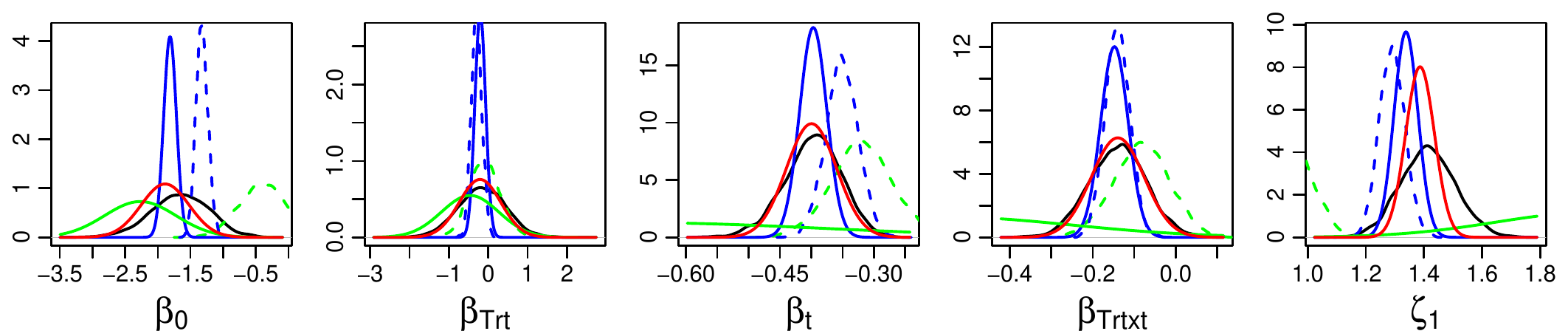}
\caption{Toenail data: posterior distributions of $\beta$ and $\zeta$ estimated using ADVI (dotted, blue for mean-field and green for unrestricted), Algorithm 1 (blue for mean-field and green for unrestricted), Algorithm 2 (red) and MCMC (black). \label{toenail_posteriors}}
\end{figure*}

Figure \ref{toenail_posteriors} shows the variational posteriors estimated by ADVI and Algorithms 1 and 2. The estimates from Algorithm 2 are closer to that of MCMC than the unrestricted and mean-field approximations of ADVI and Algorithm 1. Table \ref{runtimes} indicates that the runtime of Algorithm 1 (unrestricted) is about 1.5 times that of Algorithm 2 even though the number of iterations required is halved. 

\subsubsection{Polypharmacy  data}
The polypharmacy data set \citep{Hosmer2013} is available at \url{http://www.umass.edu/statdata/statdata/stat-logistic.html} and it contains data on 500 subjects studied over seven years. The response is whether the subject is taking drugs from 3 or more different groups. We consider the covariates, $\text{Gender}=1$ if male and 0 if female, $\text{Race}=0$ if subject is white and 1 otherwise, Age, and the following binary indicators for the number of outpatient mental health visits, MHV\textunderscore 1=1 if $1 \leq \text{MHV} \leq 5$, MHV\textunderscore 2=1 if if $6 \leq \text{MHV} \leq 14$ and MHV\textunderscore 3=1 if $\text{MHV} \geq 15$. Let INPTMHV = 0 if there were no inpatient mental health visits and 1 otherwise. We consider a logistic random intercept model \citep[see][]{Hosmer2013} of the form
\begin{equation*}\label{polymodel}
\begin{aligned}
\text{logit}(\mu_{ij})&=\beta_0+\beta_1\text{Gender}_i+\beta_2 \text{Race}_i + \beta_3 \text{Age}_{ij} \\
& + \beta_4 \text{MHV\textunderscore 1}_{ij} + \beta_5 \text{MHV\textunderscore 2}_{ij} + \beta_6 \text{MHV\textunderscore 3}_{ij} \\
& + \beta_7 \text{INPTMHV}_{ij} + u_i,
\end{aligned}
\end{equation*}
for $i=1,\dots, 500$, $j=1,\dots,7$. 

\begin{figure*}[htb!]
\centering
\includegraphics[width=\textwidth]{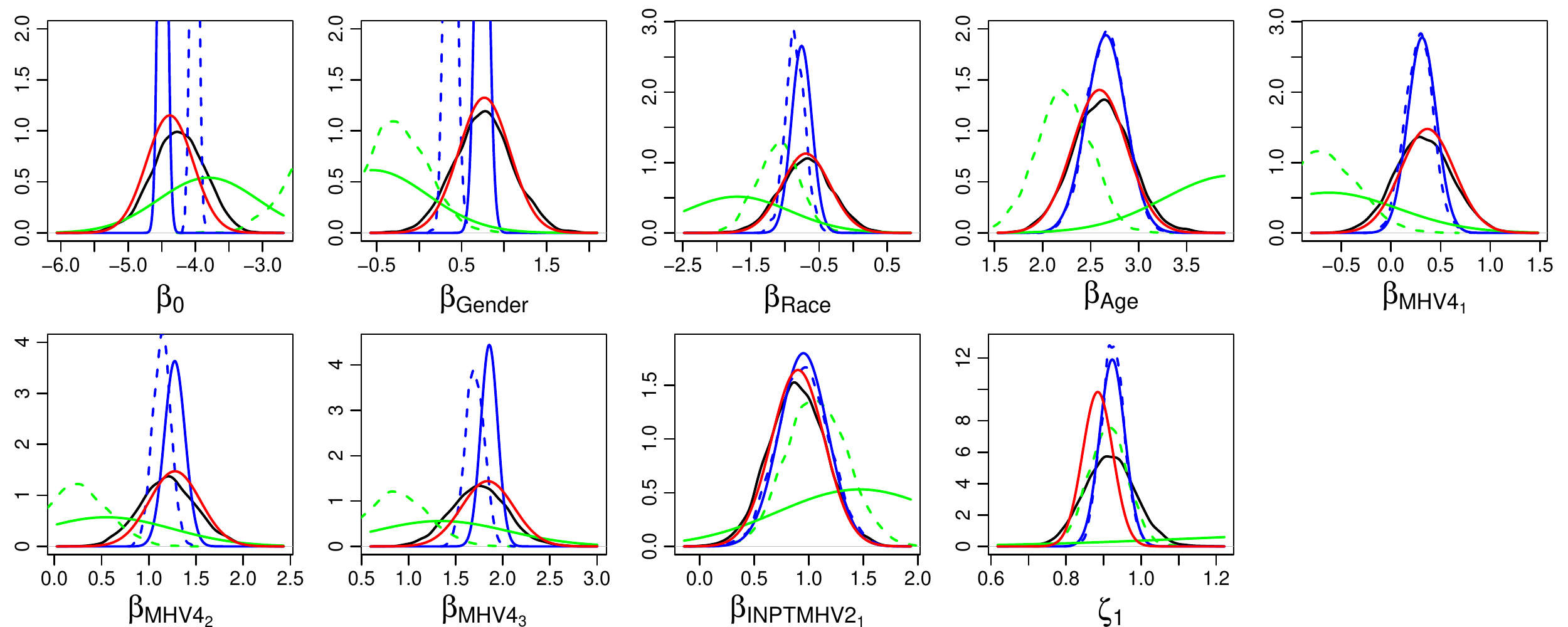}
\caption{Polypharmacy data: posterior distributions of $\beta$ and $\zeta$ estimated using ADVI (dotted, blue for mean-field and green for unrestricted), Algorithm 1 (blue for mean-field and green for unrestricted), Algorithm 2 (red) and MCMC (black) \label{polypharmacy_posteriors}.}
\end{figure*}

We apply ADVI and Algorithms 1 and 2 to this model. From Table \ref{runtimes}, the increase in runtime of Algorithm 1 (unrestricted) due to the larger number of subjects as compared to the toenail dataset is evident. The runtime of Algorithm 1 (unrestricted) is about 4.7 times that of Algorithm 2 while the runtime of Algorithm 2 is only slightly longer than that of Algorithm 1 (mean-field). From Figure \ref{polypharmacy_posteriors}, Algorithm 2 provided a very good approximation of the marginal posteriors of $\beta$ but there is some underestimation of the mean and standard deviation of $\zeta_1$.

\subsection{State space models}
Here we consider the stochastic volatility model widely used in modeling financial time series \cite[see, e.g.][]{Harvey1994, Kim1998}, which is an example of a non-linear state space model. The observations $y_t$ are generated from a zero-mean Gaussian distribution with a variance stochastically evolving over time. The unobserved log-volatility $b_t$ is modeled as an AR(1) process with Gaussian disturbances. Let
\begin{equation} \label{SV_model}
\begin{aligned}
y_t &\sim N(0, \exp(\lambda + \sigma b_t) ) \;\text{for } t = 1, \dots, n, \\
b_1 &\sim N(0,\; 1/(1-\phi^2)),  \\
b_{t+1} &\sim N(\phi b_t, 1)\;\text{for } t = 2, \dots, n,
\end{aligned}
\end{equation}
where $\lambda \in \mathbb{R}$, $\sigma>0$ and $0 < \phi <1$. In \eqref{SV_model}, $y_t$ is the mean-corrected return at time $t$ and $b_t$ is assumed to follow a stationary process with $b_1$ drawn from the stationary distribution. We transform the constrained parameters to the real space by letting $\sigma = \exp(\alpha)$ and $\phi = \frac{\exp(\psi) }{\exp(\psi) +1}$, where $\alpha, \psi \in \mathbb{R}$. Assume normal priors, $\alpha \sim N(0, \sigma_\alpha^2)$, $\lambda \sim N(0, \sigma_\lambda^2)$ and $\psi \sim N(0, \sigma_\psi^2)$. The set of variables is given by $\theta = (b_1, \dots, b_n, \eta^T)^T$, where $\eta = (\alpha, \lambda, \psi)$, and the joint distribution is given by 
\begin{equation} \label{svlogjoint}
\begin{aligned}
p(y, \theta) &= h(\theta) =p(\alpha) p(\lambda) p(\psi) p(b_1| \psi) \\
& \quad \times  \left\{\prod_{t=1}^n p(y_t|b_t, \lambda, \alpha) \right\} \left\{\prod_{t=1}^{n-1} p(b_{t-1}| b_t, \psi) \right\}.
\end{aligned}
\end{equation}
From \eqref{svlogjoint}, $b_t$ is conditionally independent of all other states in the posterior distribution given $\eta$ and the neighboring states. Hence, we can take advantage of this conditional independence in the variational approximation $q(\theta) = N(\mu,\Omega)$. By setting  $T_{ij}=0$, $1\leq j<i-1<n$, $T T^T$ has the sparsity we desire for $\Omega$. This sparse structure is illustrated in \eqref{Tstructure_SMM}.  
\begin{equation} \label{Tstructure_SMM}
\begin{small}
T = \begin{bmatrix}
T_{11} & 0 & 0 & \ldots & 0 & 0 & 0\\
T_{21} & T_{22} & 0 & \ldots & 0 & 0 & 0 \\
0 & T_{32} & T_{33} & \ldots & 0 & 0 & 0\\
\vdots & \vdots & \vdots & \ddots & \vdots & \vdots & \vdots\\
0 &  0 & 0 & \ldots & T_{n-1,n-1} & 0 & 0 \\
0 &  0 & 0 & \ldots & T_{n,n-1} & T_{n,n} & 0 \\
T_{n+1,1} & T_{n+1,2} & T_{n+1,3} & \ldots  & T_{n+1,n-1} & T_{n+1,n} & T_{n+1,n+1} \\
\end{bmatrix}
\end{small} 
\end{equation}
For the SSM, the number of parameters to update in each iteration of Algorithm 1 (unrestricted)  is $\mathcal{O}(n^2)$ while Algorithm 2 only requires $\mathcal{O}(n)$ updates (similar to Algorithm 1 mean-field). This is an important factor to consider in SSMs as the number of observations in a time series over a long period may be large. 

Next, we illustrate Algorithm 2 using two sets of exchange rate data which is available from the dataset ``Garch" in the R package {\ttfamily Ecdat}. We compute the mean-corrected response $\{ y_t \}$ from the exchange rates $\{r_t\}$ as 
\begin{equation*}
y_t = 100 \times \left\{ \log({r_t}/{r_{t-1}}) - \frac{1}{n} \sum_{i=1}^n   \log({r_i}/{r_{i-1}})  \right\}.
\end{equation*}
The gradient of $\log h(\theta)$ is derived in \ref{SSM_deriv}.

\subsubsection{GBP/USD exchange rate data}
\begin{figure*}[htb!]
\centering
\includegraphics[width=0.7\textwidth]{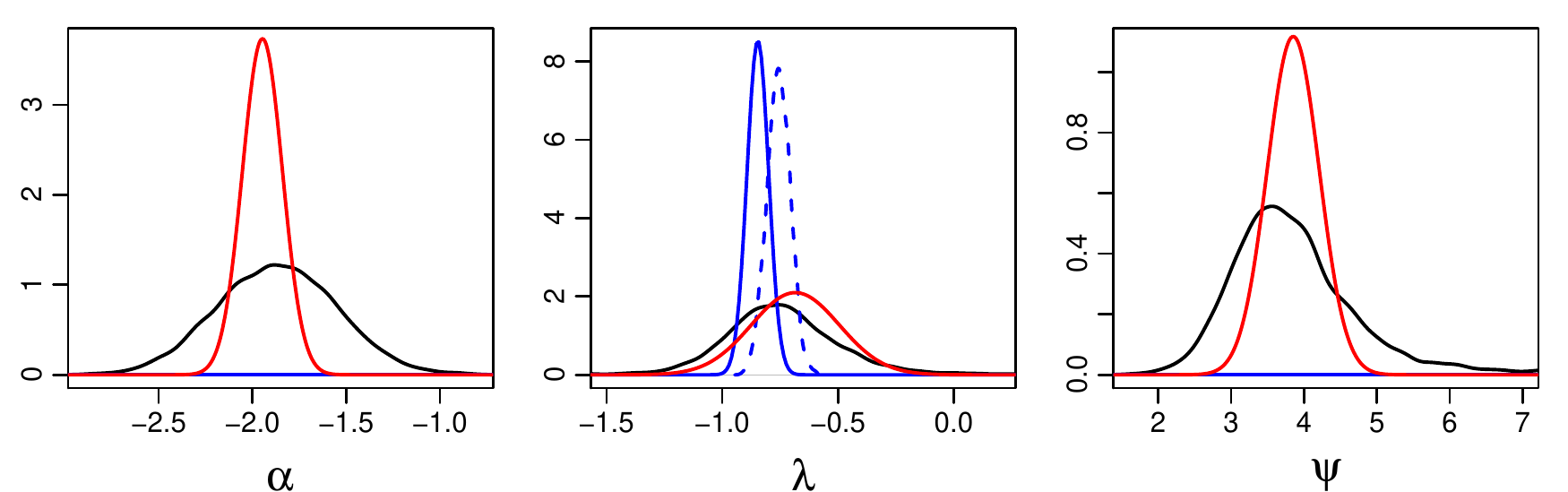}
\caption{GBP/USD exchange rate data: posterior distributions of $\{\alpha, \lambda, \psi\}$ estimated using ADVI (dotted, blue for mean-field), Algorithm 1 (blue for mean-field), Algorithm 2 (red) and MCMC (black). \label{GBP_posteriors}}
\end{figure*}

\begin{figure*}[htb!]
\centering
\includegraphics[width=0.75\textwidth]{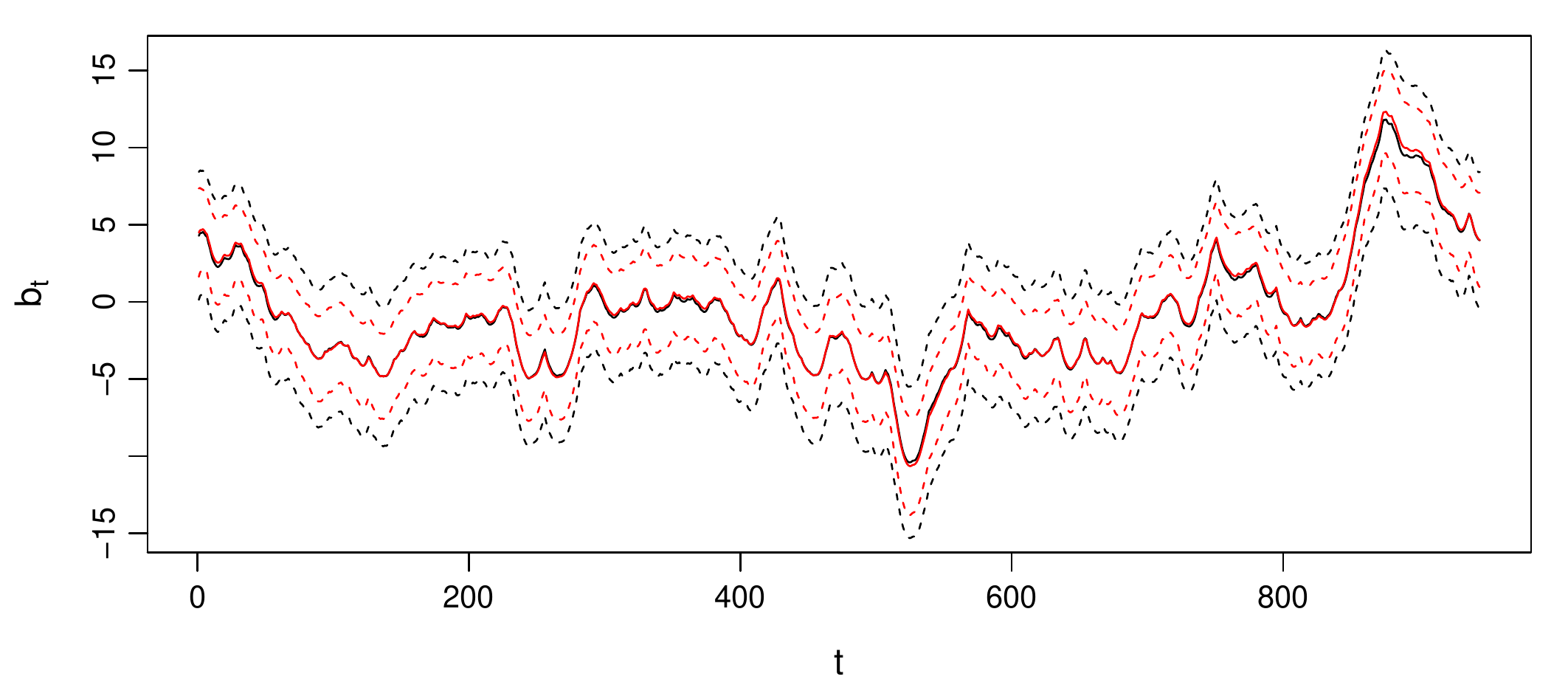}
\caption{GBP/USD exchange rate data: Mean (solid line) and 1 standard deviation intervals (dotted lines) of log volatility $b_t$ estimated using Algorithm 2 (red) and MCMC (black). \label{GBP_h}}
\end{figure*}

Here we consider daily observations of the weekday exchange rates of the U.S. Dollar against the British Pound from 1st October 1981 to 28th June 1985. This dataset has been considered by \cite{Harvey1994}, \cite{Kim1998} and \cite{Durbin2012}. The number of responses is $n=945$. Applying ADVI and  Algorithms 1 and 2 to this dataset, the resulting variational posteriors are shown in Figure \ref{GBP_posteriors}. We note that Algorithm 1 (unrestricted) diverges as the averaged lower bound $\bar{\mathcal{L}}$ is deteriorating and tending towards $-\infty$. ADVI (unrestricted) also fails to converge. The mean-field approximations of ADVI and Algorithm 1 have difficulty in capturing the means of $\alpha$ and $\psi$ and only manage to capture the mean of $\lambda$. Algorithm 2 was able to capture the means with reasonable accuracy but there is underestimation in the variance of $\alpha$ and $\psi$. 
Figure \ref{GBP_h} shows the mean (solid lines) and 1 standard deviation intervals (dotted lines) of the log volatility $b_t$ at each time point estimated using Algorithm 2 and MCMC. Algorithm 2 was able to capture the means very accurately but there is some underestimation of the standard deviation.

\subsubsection{DEM /USD exchange rate data}
\begin{figure*}[htb!]
\centering
\includegraphics[width=0.95\textwidth]{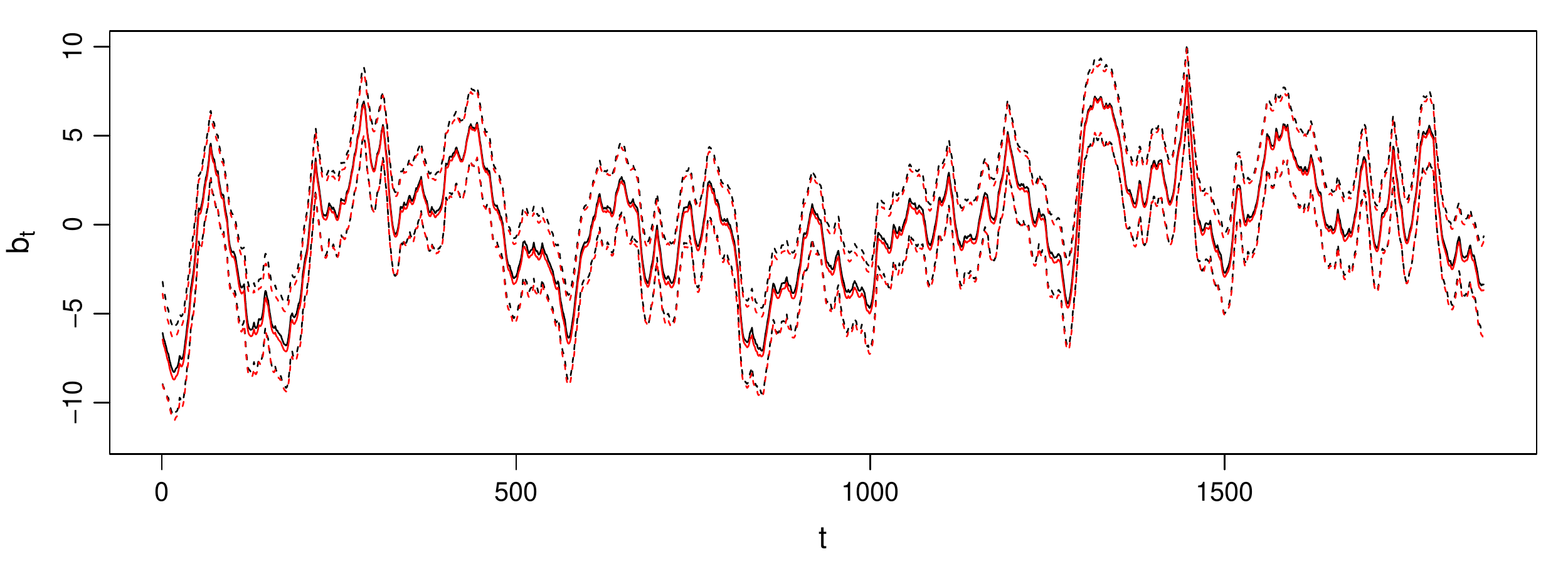}
\caption{DEM/USD exchange rate data: Mean (solid line) and 1 standard deviation intervals (dotted lines) of log volatility $b_t$ estimated using Algorithm 2 (red) and MCMC (black). \label{DEM_h}}
\end{figure*}

Next, we consider the entire series of weekday exchange rates of the U.S. Dollar against the German Deutschemark from 2nd January 1980 to 21st June 1987 available in ``Garch". This is a much larger dataset with $n=1866$ responses. We apply ADVI and Algorithms 1 and 2 to this dataset. The unrestricted approximations of ADVI and Algorithm 1 fail to converge again. The approximations of Algorithm 2 improved from the previous dataset and the underestimation of the standard deviation was less severe. As in the previous case, the mean field approximations of ADVI and Algorithm 1 had difficulty in capturing the means of $\alpha$ and $\psi$.
\begin{figure*}[htb!]
\centering
\includegraphics[width=0.7\textwidth]{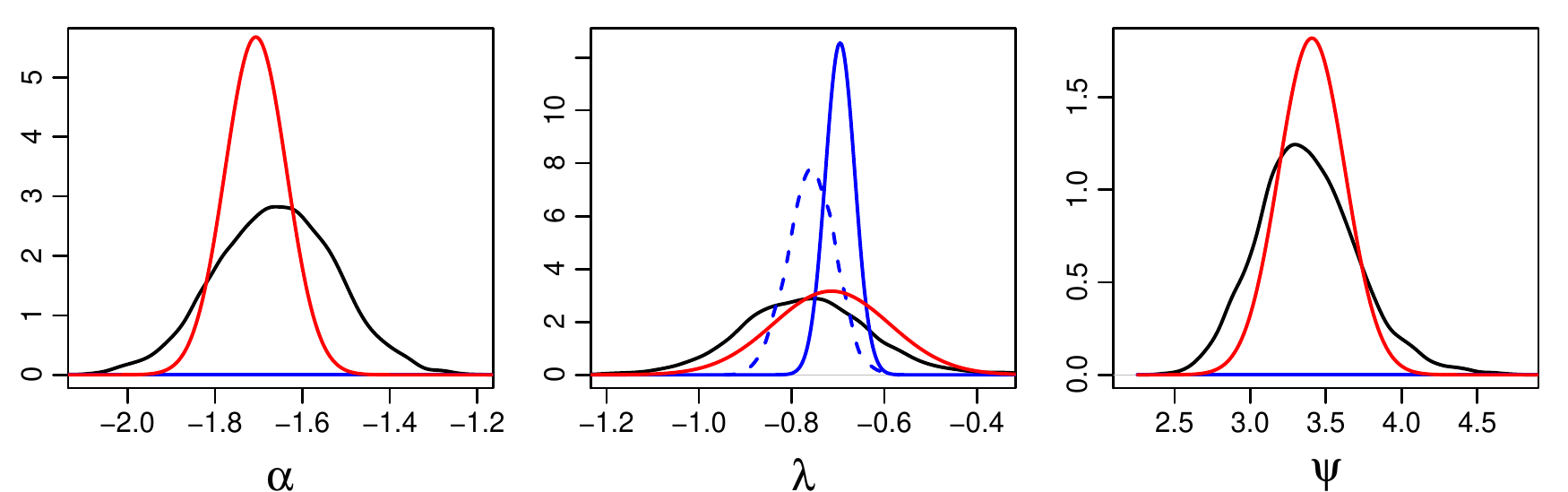}
\caption{DEM/USD exchange rate data: posterior distributions of $\{\alpha, \lambda, \psi\}$ estimated using ADVI (dotted, blue for mean-field), Algorithm 1 (blue for mean-field), Algorithm 2 (red) and MCMC (black).}
\end{figure*}
Figure \ref{DEM_h} shows the mean and 1 standard deviation intervals of the log volatility $b_t$. For this dataset, Algorithm 2 captured both the mean and standard deviation of the log volatility accurately.

\section{Conclusion} \label{sec: conclude}
In this article, we propose parameterizing the precision matrix in a Gaussian variational approximation in terms of the Cholesky factor and performing optimization using stochastic gradient methods. This approach enables us to impose sparsity in the precision matrix so as to reflect conditional independence structure in the posterior distribution appropriately. We also propose improved estimators of the stochastic gradient, which have lower variance and which are helpful in increasing the stability and precision of the algorithm. We illustrate how our approach can be applied to generalized linear mixed models and state space models. Our experimental results indicate that imposing sparsity in the precision matrix reduces the computational complexity of the problem greatly without having to assume independence among the model parameters. This also helps to improve the accuracy of the posterior approximations. We note that Algorithm 2 performs better than the unrestricted approximations of ADVI and Algorithm 1 on several occasions even though there are lesser constraints in the unrestricted approximations. This may be due to the difficulties associated with such high-dimensional optimization. In using a Gaussian variational approximation, all constrained parameters have to be transformed to the real line and we observed sensitivity towards the transformations being used as well as the way the model is being parameterized. These are important areas for future work.

\begin{acknowledgements}
Linda Tan was supported by the National University of Singapore Overseas Postdoctoral Fellowship. David Nott's research was supported by a Singapore Ministry of Education Academic Research Fund Tier 2 grant (R-155-000-143-112). We thank the reviewers and the editors for their time and helpful comments which have improved the manuscript. 
\end{acknowledgements}

\bibliographystyle{chicago}
\bibliography{ref}

\begin{appendices}
\renewcommand\thesection{Appendix \Alph{section}}

\section{Gradients for generalized linear mixed models} \label{GLMM_deriv}
For the GLMM described in Section \ref{sec:GLMM}, we have
\begin{equation*}
\begin{aligned}
\log h(\theta) & = \sum_{i=1}^n \sum_{j=1}^{n_i} \log p(y_{ij}|\beta, b_i) + \sum_{i=1}^n \log p(b_i|\zeta) \\
& \quad + \log p(\beta) + \log p(\zeta)  \\
& = \sum_{i,j} \{y_{ij}(X_{ij}^T \beta + Z_{ij}^T b_i) - h_1(X_{ij}^T \beta + Z_{ij}^T b_i)\} \\
& \quad -n\log|W|  -\frac{1}{2} \sum_{i=1}^n b_i^T W^{-T}W^{-1}b_i \\
& \quad -\frac{1}{2\sigma_\beta^2}\beta^T\beta -\frac{1}{2\sigma_\zeta^2}\zeta^T\zeta +C,
\end{aligned}
\end{equation*}
where $C$ is a constant not dependent on $\theta$. For the logistic GLMM, $h_1(x) =  \log \{1+\exp(x)\}$ while for the Poisson GLMM, $h_1(x) = \exp(x)$. The gradient of $\log h(\theta)$ is given by 
$[\nabla_{b_1} \log h(\theta), \dots, \nabla_{b_n} \log h(\theta), \nabla_{\beta} \log h(\theta), \nabla_{\zeta} \log h(\theta)] $, where
\begin{equation} \label{grad_GLMM}
\begin{aligned}
\nabla_{b_i} \log h(\theta) &= \sum_{j=1}^{n_i} \{y_{ij} - h_1'(X_{ij}^T \beta + Z_{ij}^T b_i)  \}Z_{ij} \\
& \quad - W^{-T}W^{-1}b_i, \text{ for } i=1,\dots, n, \\
\nabla_{\beta} \log h(\theta) &= \sum_{i=1}^n \sum_{j=1}^{n_i} \{y_{ij} -h_1'(X_{ij}^T \beta + Z_{ij}^T b_i) \}X_{ij} \\
& \quad - \frac{1}{\sigma_\beta^2} \beta, \\
\nabla_{\zeta} \log h(\theta) &=  -n1_{\text{diag}(W)}  + 1_\zeta \odot  \text{vech}(A)  -\frac{1}{\sigma_\zeta^2}\zeta .
\end{aligned}
\end{equation}
In \eqref{grad_GLMM}, $A= \sum_{i=1}^n W^{-T}W^{-1}  b_i b_i^T W^{-T}$ with all entries above the diagonal set to zero and $1_{\text{diag}(W)}$ and $1_\zeta$ are vectors of length $p(p+1)/2$. We define the $i$th element of $1_{\text{diag}(W)}$ as 1 if the $i$th element of $\zeta$ correspond to a diagonal element of $W$ and 0 otherwise. On the other hand, the $i$th element of $1_\zeta$ is $\exp(\zeta_i)$ if $\zeta_i$ corresponds to a diagonal element of $W$ and 1 otherwise. For the logistic GLMM, $h_1'(x) = \{ 1+\exp(-x) \}^{-1}$ and for the Poisson GLMM, $h_1'(x) = \exp(x)$. More details on the derivation of the gradients are given below. 

As $\log|W|  =  \sum_{i=1}^p W^*_{ii}$,  $\nabla_{\zeta} \log|W| = 1_{\text{diag}(W)}$. For the term $ -\frac{1}{2}\sum_{i=1}^n b_i^T W^{-T}W^{-1}b_i$, 
\begin{equation*}
\begin{aligned}
&\text{d}\, b_i^T W^{-T}W^{-1}b_i \\
&= - \{b_i^T W^{-T} (\text{d} W^T) W^{-T} W^{-1}b_i \\
& \quad  + b_i^T W^{-T} W^{-1} (\text{d} W) W^{-1}b_i\}   \\
& = - \{  (b_i^T W^{-T} W^{-1} \otimes b_i^T W^{-T}) \text{vec}(\text{d} W^T)  \\
& \quad + (b_i^T W^{-T}  \otimes b_i^T W^{-T} W^{-1}) \text{vec}(\text{d} W) \}  \\
& = - \{  (b_i^T W^{-T} W^{-1} \otimes b_i^T W^{-T}) K_p  \\
& \quad  + (b_i^T W^{-T}  \otimes b_i^T W^{-T} W^{-1})  \} \text{vec}(\text{d} W) \\
& = - \{  (b_i^T W^{-T} W^{-1} \otimes b_i^T W^{-T}) K_p  \\
& \quad + K_1(b_i^T W^{-T} W^{-1} \otimes b_i^T W^{-T}) K_p \} \text{d}\text{vec}(W) \\
& = -2 (b_i^T W^{-T} W^{-1} \otimes b_i^T W^{-T}) K_p \text{d}\text{vec}(W) \\
& = -2 \text{vec}(W^{-1}b_ib_i^T W^{-T} W^{-1})^T K_p \text{d}\text{vec}(W) \\
& = -2 \text{vec}(W^{-T}W^{-1}b_ib_i^T W^{-T} )^T \text{d}\text{vec}(W),
\end{aligned}
\end{equation*}
where $K_p$ denotes the $p^2 \times p^2$ commutation matrix. Let $A = \sum_{i=1}^n W^{-T}W^{-1}b_ib_i^T W^{-T}$ with all entries above the diagonal set to zero. As $W$ is a lower triangular matrix,  
\begin{equation*}
\nabla_{\text{vec}(W)}(-\frac{1}{2} \sum_{i=1}^n b_i^T W^{-T}W^{-1}b_i)  = \text{vec}(A).
\end{equation*}
Moreover,
\begin{equation*}
\text{d}\text{vec}(W) = D_p \text{d}\text{vech}(W) = D_p \text{diag}(1_\zeta) \text{d} \zeta,
\end{equation*}
where $D_p$ is the $p(p+1)/2 \times 1$ duplication matrix. Therefore, using chain rule
\begin{equation*}
\begin{aligned}
\nabla_{\zeta}(-\frac{1}{2} \sum_{i=1}^n b_i^T W^{-T}W^{-1}b_i) &=  \text{diag}(1_\zeta) D_p^T\text{vec}(A)  \\
& = 1_\zeta \odot  \text{vech}(A +A^T - \text{dg}A) \\
& = 1_\zeta \odot  \text{vech}(A),
\end{aligned}
\end{equation*}
where $\text{dg}A$ is a diagonal matrix with diagonal equal to the diagonal of A. The last line follows because $A$ is a lower triangular matrix.

\section{Gradients for state space model} \label{SSM_deriv}
For the stochastic volatility model in \eqref{svlogjoint}, 
\begin{equation*}
\begin{aligned}
\log h(\theta) &= -\frac{n\lambda}{2}   - \frac{\e^\alpha}{2} \sum_{t=1}^n b_t  -  \frac{1}{2} \sum_{t=1}^n y_t^2\exp(-\lambda - \e^\alpha b_t) \\
& \quad  - \frac{1}{2} \sum_{t=1}^{n-1} (b_{t+1} - \phi b_t)^2 + \frac{1}{2} \log(1-\phi^2) \\
& \quad  - \frac{1}{2} (1-\phi^2)b_1^2 - \frac{\alpha^2}{2\sigma_\alpha^2} - \frac{\lambda^2}{2\sigma_\lambda^2} - \frac{\psi^2}{2\sigma_\psi^2} +C,
\end{aligned}
\end{equation*}
where $C$ is a constant independent of $\theta$.
The gradient $\nabla_{\theta} \log h(\theta)$ can be computed from the following components.
\begin{equation*}
\begin{aligned}
\nabla_{b_1} \log h(\theta) &= - (1-\phi^2) b_1 + \phi (h_2 - \phi b_1)  -\frac{\e^\alpha}{2}  \\
& \quad +  \frac{ \e^\alpha}{2} y_1^2\exp(-\lambda - \e^\alpha b_1), \\
\nabla_{b_t} \log h(\theta) &= \phi (b_{t+1} - \phi b_t) - (b_t - \phi b_{t-1}) -\frac{\e^\alpha}{2} \\
& \quad +  \frac{ \e^\alpha}{2} y_t^2\exp(-\lambda - \e^\alpha b_t) \text{ for }1 < t < n,  \\
\nabla_{b_n} \log h(\theta) &= - (b_n - \phi h_{n-1}) -\frac{\e^\alpha}{2}  \\
& \quad  +  \frac{ \e^\alpha}{2} y_n^2\exp(-\lambda - \e^\alpha b_n), \\
\nabla_{\alpha} \log h(\theta) &=  \frac{1}{2} \sum_{t=1}^n y_t^ 2b_t \exp(\alpha -\lambda - \e^\alpha b_t) \\
& \quad - \frac{\e^\alpha}{2} \sum_{t=1}^n b_t  - \frac{\alpha}{\sigma_\alpha^2} , \\
\nabla_{\lambda} \log h(\theta) &=  -\frac{n}{2}  +  \frac{1}{2} \sum_{t=1}^n y_t^2\exp(-\lambda - \e^\alpha b_t)  - \frac{\lambda}{\sigma_\lambda^2}, \\
\nabla_{\psi} \log h(\theta) &= \Big\{ \phi b_1^2- \frac{\phi}{(1-\phi^2)} +  \sum_{t=1}^{n-1} (b_{t+1} - \phi b_t)h_t  \Big\}\\
& \quad \times\frac{e^\psi}{(e^\psi + 1)^2}  - \frac{\psi}{\sigma_\psi^2}. \\
\end{aligned}
\end{equation*}

\end{appendices}

\end{document}